***version for ArXiv***

# Symmetry of flexoelectric response in ferroics

Eugene A. Eliseev[1], Anna N. Morozovska[2,3*], Victoria V. Khist[4], and Victor Polinger[5†]

[1] *Institute for Problems of Materials Science, National Academy of Sciences of Ukraine, 3, Krjijanovskogo, 03142 Kyiv, Ukraine*

[2] *Institute of Physics, National Academy of Sciences of Ukraine, 46, Prospekt Nauky, 03028 Kyiv, Ukraine*

[3] *Bogolyubov Institute for Theoretical Physics, National Academy of Sciences of Ukraine, 14-b Metrolohichna, 03680 Kyiv, Ukraine*

[4] *Institute of Magnetism, National Academy of Sciences of Ukraine and Ministry of Education and Science of Ukraine, Prospekt Vernadskogo 36a, 03142 Kyiv, Ukraine*

[5] *Department of Chemistry, University of Washington, 109 Bagley Hall, Seattle, WA 98195, USA*

## Abstract

Using direct matrix method we establish the structure, including the number of nonzero independent elements, of the static flexoelectric coupling tensor $f_{ijkl}$ for all 32 point groups. We use the point symmetry of elementary cell, previously known evident index-permutation symmetry ( $f_{ijkl} \equiv f_{jikl}$ ) and recently established "hidden" index-permutation symmetry ( $f_{ijkl} = f_{ilkj}$ ). We compare these results and demonstrated that the hidden symmetry of $f_{ijkl}$ significantly reduces the number of its nonzero independent elements. Using group theory, we find out the explicit form of the tensor $f_{ijkl}$ and the flexoelectric coupling energy in the form of Lifshitz invariant, $\frac{f_{ijkl}}{2}\left(P_k\frac{\partial u_{ij}}{\partial x_l} - u_{ij}\frac{\partial P_k}{\partial x_l}\right)$, for several point symmetries most important for applications. For these symmetries we visualize the effective flexoelectric response of the bended plate allowing for both evident and hidden index-permutation symmetries, analyze its anisotropy, and angular dependences. We discuss how the hidden symmetry can significantly simplify the problem of the flexoelectric constants determination from inelastic neutron and Raman scattering experiments due to the minimization of $f_{ijkl}$ independent components, opening the way for its unambiguous determination from the scattering experiments.

---

* anna.n.morozovska@gmail.com

† VPolinger@msn.com

# I. INTRODUCTION

**A. Index-permutation symmetry of the flexoelectric coupling.** The static flexoelectric effect is the appearance of elastic strain $u_{ij}$ in response to electric polarization gradient $\partial P_k/\partial x_l$ (direct effect), and, vice versa, the polarization $P_i$ appears as a response to the strain gradient $\partial u_{ij}/\partial x_l$ (inverse effect) [1, 2, 3, 4]. In continuum media approximation the static flexoelectric coupling (for brevity "**flexocoupling**") is included to the free energy functional of a ferroic in the form of Lifshitz invariant [5, 6, 7, 8],

$$F_{FL} = \int_V d^3 r \frac{f_{ijkl}}{2}\left(P_k \frac{\partial u_{ij}}{\partial x_l} - u_{ij}\frac{\partial P_k}{\partial x_l}\right). \tag{1}$$

Here the strain tensor components $u_{ij} = \left(\partial U_i/\partial x_j + \partial U_j/\partial x_i\right)/2$ are symmetrized derivatives of the displacement vector components $U_i$. Notably that the antisymmetric part of the strain tensor does not contribute to the flexoelectric response of the bulk material being a full derivative [8], and the form (1) is the most comprehensive one. Due to the evident index-permutation symmetry of the strain tensor, $u_{ij} \equiv u_{ji}$, all components of the static flexocoupling tensor $f_{ijkl}$ are symmetrical (i.e. index-permutative) with respect to the permutation of the first pair of indices,

$$f_{ijkl} \equiv f_{jikl}. \tag{2a}$$

The static flexoelectric effect is allowed by symmetry in all 32 crystalline point groups, because the strain gradient breaks the inversion symmetry giving rise to the static flexoelectricity. For instance, Shu et al. [9], Quang and He [10], consider possible symmetries of the flexoelectric tensor and derived the number of its independent components for each symmetry. Let us underline the fact that Shu et al.[9], and Quang and He [10] used the invariance of the flexocoupling tensor $f_{ijkl}$ to the permutation of the first one pair of indices, Eq.(2a).

In the case of an infinite continuous medium, one can apply integration by parts to Eq.(1). In addition to the "**evident**" index-permutation symmetry (2a), this reveals more, previously "**hidden**" index-permutation symmetry properties [11], namely:

$$f_{ijkl} = f_{ilkj} \tag{2b}$$

Hence both the index-permutation equations (2) are valid for the static flexocoupling tensor $f_{ijkl}$.

Using the symmetry theory, Kvasov and Tagantsev [12] predicted the existence of dynamic flexoelectric effect. They argued that dynamic equations of state for a condensed media can be derived from the minimization of Lagrange function, $L = F - T$, where the free energy $F$ is given by expression

$$F = \int_V d^3 r \left( \alpha P_i P_i + \alpha_{ijkl} P_i P_j P_k P_l + \frac{g_{ijkl}}{2}\left(\frac{\partial P_i}{\partial x_j}\frac{\partial P_k}{\partial x_l}\right) - q_{ijkl} u_{ij} P_k P_l + \frac{c_{ijkl}}{2} u_{ij} u_{kl} - P_i E_i - N_{ij} u_{ij} \right) \quad (3)$$
$$+ F_{FL}$$

Hereinafter the summation is over all repeating indexes; the coefficient $\alpha$ is temperature dependent for ferroics, $q_{mnij}$ is electrostriction tensor, the higher-order coefficients $\alpha_{ijkl}$ are regarded temperature independent, $g_{ijkl}$ are gradient coefficients tensor**,** $c_{ijkl}$ are elastic compliances. $N_{ij}$ is the anisotropic external load, $E_i$ is electric field that obeys electrostatic equation, $\varepsilon_b \varepsilon_0 \frac{\partial E_i}{\partial x_i} = -\frac{\partial P_i}{\partial x_i}$ ($\varepsilon_b$ is background permittivity [13] and $\varepsilon_0 = 8.85\times 10^{-12}$ F/m is the vacuum dielectric constant). The kinetic energy $T$ is given by

$$T = \frac{\mu}{2}\left(\frac{\partial P_i}{\partial t}\right)^2 + M_{ij}\frac{\partial P_i}{\partial t}\frac{\partial U_j}{\partial t} + \frac{\rho}{2}\left(\frac{\partial U_i}{\partial t}\right)^2, \quad (4)$$

which includes the dynamic flexoelectric coupling that is described by the second rank tensor $M_{ij}$ [12]. $U_i$ is elastic displacement, $\mu$ is a dynamic coefficient and $\rho$ is the density of a ferroelectric.

**B. The work motivation, novelty and content.** Despite the index-permutation symmetry (2b) has been recently established, its impact on the explicit form of the flexocoupling tensor for all 32 point groups is absent.

Using direct method, described in Ref. [14] and applied there to the flexomagnetic tensor, here we studied the structure of $f_{ijkl}$, including the number of nonzero independent elements for all 32 groups . At first we use point symmetry and evident index-permutation symmetry (2a) only (see **Table AI** in **Appendix A1** in the **Supplement** [15]). Next we apply point symmetry, index-permutation symmetry (2a) and hidden index-permutation symmetry (2b) altogether (see **Table AII** in **Appendix A1** in the **Supplement** [15]). Then we performed the comparison of these results and demonstrated that the application of index-permutation symmetry (2b) significantly reduces the number of independent elements of $f_{ijkl}$ (see **Table AIII** in **Appendix A1** in the **Supplement** [15]). These results are resumed in **Section II.** Also we derive the explicit form of the tensor $f_{ijkl}$ and the Lifshitz invariant (1) for several point symmetries using the direct method for the invariants derivation and group-theoretical approach (see **Appendixes A2** and **B** in the **Supplement** [15]). We visualize the effective flexoelectric response of a cubic and tetragonal ferroic plate in bending experiment allowing for the index-permutation symmetry (2) of $f_{ijkl}$ in **Section III**. Further we discuss the problem of the flexocoupling constants determination from inelastic neutron and Raman scattering experiments [16, 17, 18, 19, 20, 21, 22, 23] in **Section IV**. The special attention is drawn to the role of the index-permutation symmetry (2b) that can significantly simplify the fitting procedure due to the reduction of $f_{ijkl}$ independent components, and consequently opening the way for its unambiguous determination from the scattering experiments. **Section V** is a brief summary.

## II. FLEXOELECTRIC COUPLING IN FERROICS WITH CUBIC AND LOWER SYMMETRY

Direct application of point group symmetry operations to the flexoelectric tensor $f_{ijkl}$ (shortly "**direct method**" [24]) along with the index-permutation symmetry (2a), leads to the system of linear algebraic equations:

$$\begin{cases} f_{ijkl} = C_{ii'}C_{jj'}C_{kk'}C_{ll'}f_{i'j'k'l'}, \\ f_{ijkl} = f_{jikl}. \end{cases} \tag{5a}$$

The first line in Eq.(5a) includes point group symmetry, where matrix elements $C_{ii'}$ are concrete point group symmetry operations in the matrix form, and the second line reflects the index-permutation symmetry (2a). Solution of Eq.(5a) gives the detailed structure of $f_{ijkl}$ for all 32 point symmetry groups, including the number of nonzero and independent elements, (see **Table AI** in **Appendix A1** [15], and compare it with the results of Shu et al.[9], and Quang and He [10]).

More complex the system of linear algebraic equations,

$$\begin{cases} f_{ijkl} = C_{ii'}C_{jj'}C_{kk'}C_{ll'}f_{i'j'k'l'}, \\ f_{ijkl} = f_{jikl}, \\ f_{ijkl} = f_{ilkj} \end{cases} \tag{5b}$$

includes point group symmetry (the first line), evident index-permutation symmetry (2a) (the second line) and recently established hidden index-permutation symmetry (2b) (the third line). Solution of Eq.(5b) gives the detailed structure of $f_{ijkl}$ for all 32 point symmetry groups (see **Table AII** in **Appendix A1** [15]).

Next, we performed the comparison of the $f_{ijkl}$ obtained from Eqs.(5a) and Eq.(5b), allowing for and without hidden index-permutation symmetry application, and demonstrated that the results calculated from Eq. (5b) contains significantly smaller number of independent $f_{ijkl}$. At that the lower is the point symmetry group the bigger is the difference between the results from Eqs. (5a) and (5b) (see comparative **Table AIII** in **Appendix A1** [15]). Note, that the amount of nonzero elements is defined by both point and index-permutation symmetries as anticipated.

The sense of the results [15] can be imagined from the comparison of the **Tables I** and **II**, presented below for several point symmetries. Namely for a cubic point symmetry (corresponding to the paraelectric phase of most perovskites) application of the index-permutation symmetry (2b) reduces the number of $f_{ijkl}$ independent elements from 3 to 2, and for the tetragonal point symmetry (corresponding to the polar long-range ordered phase of many ferroelectric perovskites), it reduces from 8 to 5 independent elements. For rather widely encountered orthorhombic point symmetry the index-permutation symmetry (2b) reduces the number of $f_{ijkl}$ independent elements from 15 to 9. The amount of nonzero elements is 21 for all considered examples, but it can be different for other groups (see comparative **Table AIII** in **Appendix A1** [15]).

**Table I.** Structure of the static flexoelectric effect tensor calculated from Eqs.(5a) for several point symmetry groups

| **Point symmetry** | **Number of nonzero elements** | **Number of independent elements** | **Nonzero elements and relations between them** |
|---|---|---|---|
| **cubic**<br>m3m, 432,<br>4'3m | 21 | 3<br>(without "hidden" index-permutation symmetry) | $f_{1111}=f_{2222}=f_{3333}$,<br>$f_{1122}=f_{2211}=f_{1133}=f_{3311}=f_{2233}=f_{3322}$,<br>$f_{1221}=f_{2112}=f_{1331}=f_{3113}=f_{2332}=f_{3223}=f_{1212}=f_{2121}=f_{1313}=f_{3131}=f_{2323}=f_{3232}$. |
| **tetragonal**<br>4'2m, 422,<br>4mm, 4/mmm | 21 | 8<br>(without "hidden" index-permutation symmetry) | $f_{3333}$, $f_{1111}=f_{2222}$, $f_{1122}=f_{2211}$,<br>$f_{1133}=f_{2233}$, $f_{3311}=f_{3322}$,<br>$f_{1212}=f_{2112}=f_{1221}=f_{2121}$,<br>$f_{1313}=f_{3113}=f_{2323}=f_{3223}$,<br>$f_{1331}=f_{3131}=f_{2332}=f_{3232}$. |
| **orthorhombic**<br>222, mm2,<br>mmm | 21 | 15<br>(without "hidden" index-permutation symmetry) | $f_{1111}$, $f_{1122}$, $f_{1133}$, $f_{2211}$,<br>$f_{2222}$, $f_{2233}$, $f_{3311}$, $f_{3322}$, $f_{3333}$,<br>$f_{1212}=f_{2112}$, $f_{1221}=f_{2121}$, $f_{1313}=f_{3113}$,<br>$f_{1331}=f_{3131}$, $f_{2323}=f_{3223}$, $f_{2332}=f_{3232}$. |

**Table II.** Structure of the static flexoelectric effect tensor calculated from Eqs.(5b) for several point symmetry groups

| **Point symmetry** | **Number of nonzero elements** | **Number of independent elements** | **Nonzero elements and relations between them** |
|---|---|---|---|
| **cubic**[*] m3m, 432, 4'3m | 21 | 2 ("hidden" index-permutation symmetry is included) | $f_{1111} = f_{2222} = f_{3333}$, $f_{1122} = f_{2211} = f_{1133} = f_{3311} = f_{2233} = f_{3322} = f_{1221} = f_{2112} = f_{1331} = f_{3113} = f_{2332} = f_{3223} = f_{1212} = f_{2121} = f_{1313} = f_{3131} = f_{2323} = f_{3232}$ |
| **tetragonal** 4'2m, 422, 4mm, 4/mmm | 21 | 5 ("hidden" index-permutation symmetry is included) | $f_{3333}$, $f_{1111} = f_{2222}$, $f_{1122} = f_{1221} = f_{2121} = f_{2211} = f_{2112} = f_{1212}$, $f_{1133} = f_{1331} = f_{3131} = f_{2233} = f_{3232} = f_{2332}$, $f_{1313} = f_{3311} = f_{3113} = f_{2323} = f_{3322} = f_{3223}$ |
| **orthorhombic** 222, mm2, mmm | 21 | 9 ("hidden" index-permutation symmetry is included) | $f_{1111}$, $f_{2222}$, $f_{3333}$ $f_{1122} = f_{1221} = f_{2121}$, $f_{2211} = f_{2112} = f_{1212}$, $f_{2233} = f_{2332} = f_{3232}$, $f_{3311} = f_{1313} = f_{3113}$, $f_{3322} = f_{3223} = f_{2323}$, $f_{1133} = f_{1331} = f_{3131}$ |

[*] there are other cubic and tetragonal point groups, not listed in **Tables AI-AII**, but listed in **Appendix A1.**

Also we derive the explicit form of the tensor $f_{ijkl}$ and the Lifshitz invariant (1) for several cubic, tetragonal and orthorhombic point symmetries using the direct method for the invariants derivation and group-theoretical approach (see **Appendixes A2** and **B** in the **Supplement** [15]).

## III. EFFECTIVE FLEXORESPONSE

The most reliable example of experimental determination of $f_{ijkl}$ is an incipient paraelectric $SrTiO_3$ having m3m symmetry (see Refs. [25, 26]). Another case is a ferroelectric family $Ba_xSr_{(1-x)}TiO_3$ ($x^3$ 0.5) having m3m parent symmetry and 4mm symmetry in a polar tetragonal phase, for which the flexoelectric coefficients have been calculated from the first principles in Refs. [27, 28]. Below we visualize the effective flexoelectric response $f_{ij}^{eff}$ (shortly **"flexoresponse"**) for $Ba_xSr_{(1-x)}TiO_3$ (x=0, 0.5, 1), using experimental results [26], *ab initio* calculations [27, 28] and elastic constants $c_{ij}$ determined experimentally in Ref.[29] and [30], respectively.

The sample orientation and the experimental setup typical for three-knife load measurements [25-26] of effective flexoresponse are shown in **Fig. 1.**

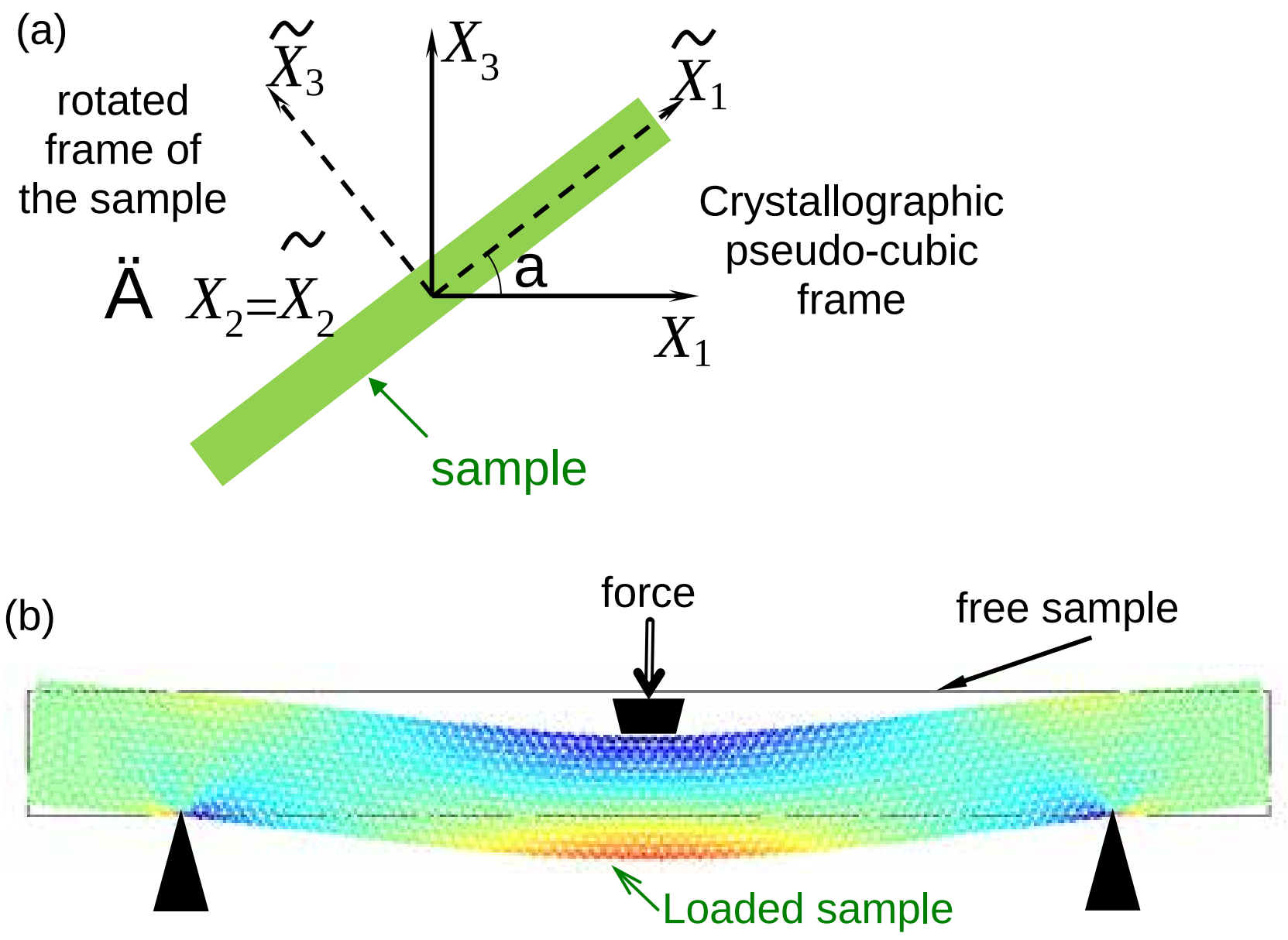

**FIGURE 1. (a)** Coordinate transformation from crystallographic (pseudo-cubic) frame $X_1, X_2, X_3$ by the rotation around $X_2$ axis on angle $\alpha$ with respect to the sample-related frame $\tilde{X}_1, \tilde{X}_2, \tilde{X}_3$. (the axis $X_2 \equiv \tilde{X}_2$ is directed perpendicular to the figure plane and shown by a crossed circle "$\otimes$"). **(b)** Typical three-knife experiment of the plate bending. Black solid lines represent cross-section of the initial plate position, rainbow-colored shape represents the deformed shape and strain distributions in the loaded plate.

Effective flexoresponse (that is a measurable value) of the plate with the surface inclined to crystallographic axes at angle $\alpha$ have two components:

$$f_{13}^{eff} = \tilde{f}_{1133} - \frac{\tilde{c}_{1133}}{\tilde{c}_{1313}\tilde{c}_{3333} - \tilde{c}_{1333}^2}\left(\tilde{c}_{1313}\tilde{f}_{3333} - \tilde{c}_{1333}\tilde{f}_{1333}\right) \tag{6a}$$

$$f_{23}^{eff} = \tilde{f}_{2233} - \frac{\tilde{c}_{2233}}{\tilde{c}_{1313}\tilde{c}_{3333} - \tilde{c}_{1333}^2}\left(\tilde{c}_{1313}\tilde{f}_{3333} - \tilde{c}_{1333}\tilde{f}_{1333}\right) \tag{6b}$$

Here we introduced the tensor components in the rotated frame, linked to the sample as shown in **Fig. 1(a)**. For the derivation of Eqs.(6) see **Appendix C** [15]. Hereinafter $\tilde{f}_{ijkl}$ and $\tilde{c}_{ijkl}$ are the "rotated" tensors of flexoelectric coefficients and elastic stiffness, respectively.

For **cubic (m3m, 432, 4'3m) groups** the "rotated" tensors $\tilde{f}_{ijkl}$ and $\tilde{c}_{ijkl}$ have the following dependence on the rotation angle $\alpha$:

$$\tilde{f}_{1133} = f_{1122} + \frac{\sin^2(2\alpha)}{2}\Delta f_{11}, \qquad \tilde{f}_{1333} = \frac{\sin(4\alpha)}{2}\Delta f_{11}, \tag{7a}$$

$$\tilde{f}_{3333} = f_{1111} - \frac{\sin^2(2\alpha)}{2}\Delta f_{11}, \quad \tilde{f}_{2233} = f_{1122}, \tag{7b}$$

$$\tilde{c}_{1133} = c_{1133} + \frac{\sin^2(2\alpha)}{2}\Delta c_{11}, \quad \tilde{c}_{1313} = c_{1313} + \frac{\sin^2(2\alpha)}{2}\Delta c_{11}, \tag{8a}$$

$$\tilde{c}_{1333} = \frac{\sin(4\alpha)}{2}\Delta c_{11}, \quad \tilde{c}_{2233} = c_{1122}, \quad \tilde{c}_{3333} = c_{1111} - \frac{\sin^2(2\alpha)}{2}\Delta c_{11}. \tag{8b}$$

Here $f_{ijkl}$ and $c_{ijkl}$ are the components of flexoelectric and elastic stiffness tensors in a crystallographic frame. In Eqs.(7)-(8) we also introduced anisotropy factors of the flexoelectric and elastic stiffness

$$\Delta f_{11} = f_{1111} - f_{3311} - f_{1313} - f_{1331}, \tag{9a}$$

$$\Delta c_{11} = c_{1111} - c_{1133} - 2c_{1313}. \tag{9b}$$

For the case of Eq.(2b) application, the expressions (9a) simplify, $\Delta f_{11} = f_{1111} - 3f_{1133}$, while Eqs.(7)-(8) remain unaffected by the hidden symmetry.

Calculated for several **cubic (m3m, 432, 4'3m)** symmetries, angular dependence of the effective flexoresponse is shown in **Fig.2.** Listed in **Table III** for the case of a thin plate made of $SrTiO_3$, the corresponding components are $f_{1122}$ = 2.64 V and $f_{1212}$ = 2.185 V. Distinguished from our predictions listed in **Tables I** and **II**, these values are not equal, $f_{1122} \neq f_{1212}$, so $f_{1122} - f_{1212}$ = 0.455 V. This difference is due to the data collected for a thin plate, finite in all 3 dimensions, where the conditions for the hidden index-permutation symmetry cannot be applied. **Fig. 2** shows angular dependences of the flexoresponse of the cubic perovskite $SrTiO_3$ with the flexoelectric constants $f_{ij}^{eff}(\alpha)$ according to Eqs.(6) - (9) and (5a). In **Fig. 2(a)**, we use experimental values of flexoelectric constants measured by Zubko [26] and listed in **Table III**, namely, $f_{1111}$ = 0.1 V, $f_{1122}$ = 2.64 V, and $f_{1212}$ = 2.185 V. The inequality $f_{1122} \neq f_{1212}$, means no index-permutation symmetry condition (2b) applies. As the effective flexoresponse was measured for a thin plate, this inequality can be interpreted as manifestation of the plate confinement in the perpendicular direction $\tilde{X}_3$.

**Fig.2(b)** shows the flexoresponse as it is expected to be in the bulk of an infinite $SrTiO_3$. The corresponding angular dependences $f_{ij}^{eff}(\alpha)$ result from the same Eqs.(6)-(9) and (5b) as above, but this time under assumption of the hidden symmetry when the index-permutation rule (2b) applies [11]. The latter requires $f_{1122} = f_{1212}$, and here we use the average off-diagonal values $f_{1122} = f_{1212}$ = 2.413 V with the same diagonal value of $f_{1111}$ = 0.1 V. Comparing solid curves in **Figs. 2(a)**, and **2(b)**, the flexoelectric component $f_{13}^{eff}$ is found to be significantly affected by the hidden symmetry (in bulk, all of its 8 lobes are of almost the same length), while there is no any significant change in the flexoelectric component $f_{23}^{eff}$ shown by the dashed curve. To summarize, the distinctive difference of **Fig. 2(a)** for a plate from **2(b)** in bulk is manifestation of the confinement effect of the thin plate in the perpendicular direction $\tilde{X}_3$.

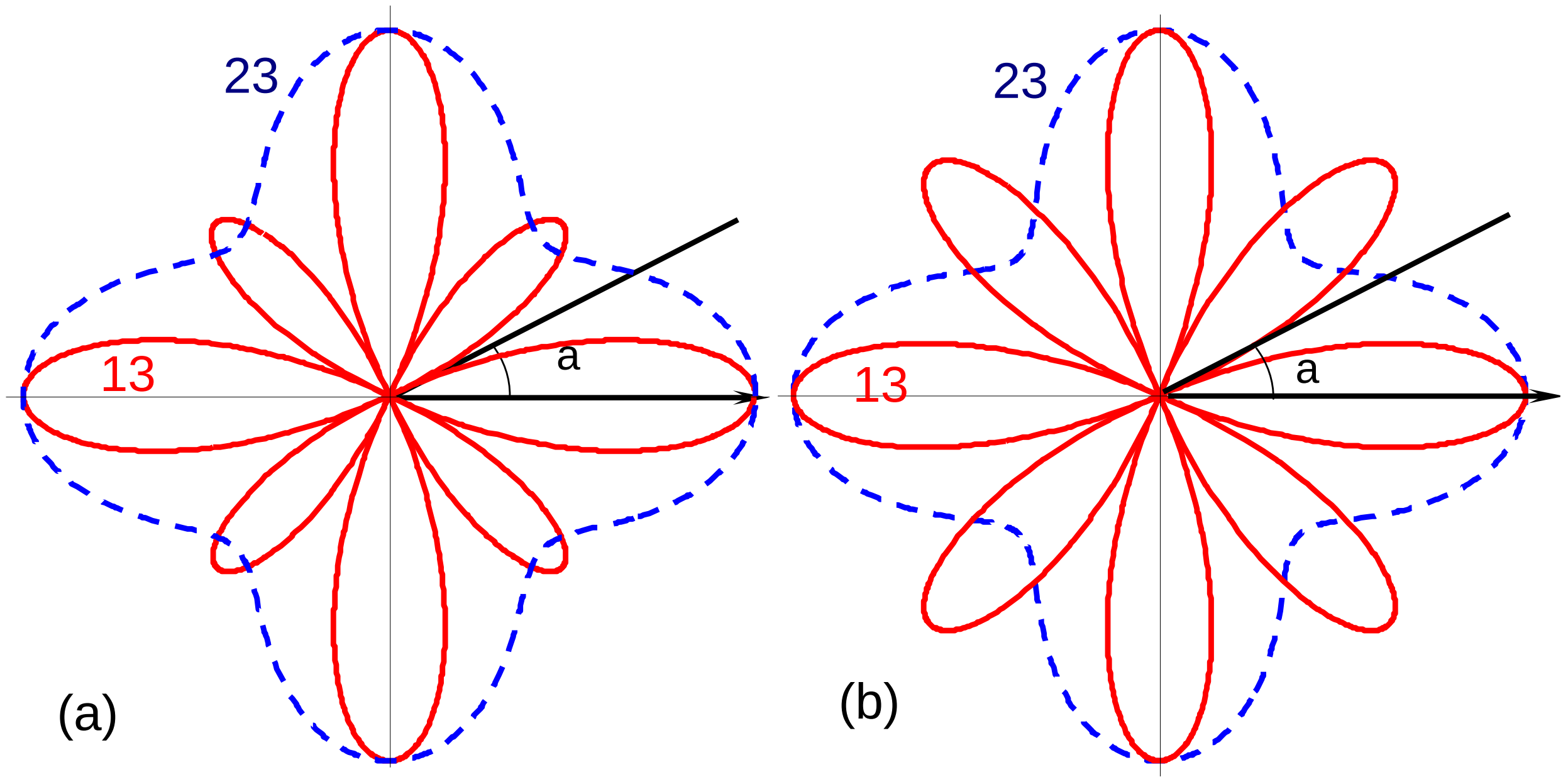

**FIGURE 2.** Angular dependence of the effective flexoresponse in a cubic crystal, symmetry *m*3*m*, for **(a)** a thin plate of $SrTiO_3$ and **(b)** in bulk of an infinite crystal $SrTiO_3$. The angle $\alpha$ represents rotation of the coordinate frame with respect to the plate-related crystallographic pseudo-cubic axis $\tilde{X}_2$ º $X_2$ (see **Fig. 1**). Solid and dashed curves represent the flexoresponse in two perpendicular directions, $f_{13}^{eff}$ and $f_{23}^{eff}$, respectively. The corresponding values of $f_{ijkl}$ and $c_{ijkl}$ are from the **Table III.** The plot **(a)** is calculated from Eqs.(5a). For the plot **(b)**, calculated from Eqs.(5b), we use symmetrized off-diagonal flexoelectric coefficients $f_{1122} = f_{1212} = 2.413$ V.

**Table III.** Flexoelectric ($f_{ijkl}$) and elastic stiffness ($c_{ijkl}$) tensor components of $SrTiO_3$

| Tensor | Numerical values | | | Reference & Notation |
|---|---|---|---|---|
| $f_{ijkl}$ (V) | $f_{1111}$= - 3.39 | $f_{1122}$=1.51 | $f_{1212}$=1.13 | *Zubko et al [25], experiment at RT |
| $f_{ijkl}$ (V) | $f_{1111}$=0.1 | $f_{1122}$=2.64 | $f_{1212}$=2.185 | **Zubko et al [26], experiment at RT |
| $c_{ijkl}$ ($10^{11}$ Pa) | $c_{1111}$=3.19 | $c_{1122}$=1.03 | $c_{1212}$=1.24 | Bell and Rupprecht [29], experiment at RT |

RT - room temperature

* initial values, obtained from incorrect interpretation of experimental measurements;

** corrected values measured experimentally.

For **tetragonal (4'2m, 422, 4mm, 4/mmm)** symmetry the "rotated" tensors $\tilde{f}_{ijkl}$ and $\tilde{c}_{ijkl}$ have the following dependence on the plate rotation angle $\alpha$:

$$\tilde{f}_{1133} = \cos^2(\alpha) f_{1133} + \sin^2(\alpha) f_{3311} + \frac{\sin^2(2\alpha)}{4}(\Delta f_{11} + \Delta f_{33}), \tag{10a}$$

$$\tilde{f}_{1333} = \frac{\sin(2\alpha)}{2}\left(\cos^2(\alpha)\Delta f_{33} - \sin^2(\alpha)\Delta f_{11}\right), \tag{10b}$$

$$\tilde{f}_{3333} = \cos^2(\alpha) f_{3333} + \sin^2(\alpha) f_{1111} - \frac{\sin^2(2\alpha)}{4}(\Delta f_{11} + \Delta f_{33}), \tag{10c}$$

$$\tilde{f}_{2233} = \sin^2(\alpha) f_{1122} + \cos^2(\alpha) f_{1133}, \tag{10d}$$

$$\tilde{c}_{1133} = c_{1133} + \frac{\sin^2(2\alpha)}{4}(\Delta c_{11} + \Delta c_{33}), \tag{11a}$$

$$\tilde{c}_{1313} = c_{1313} + \frac{\sin^2(2\alpha)}{4}(\Delta c_{11} + \Delta c_{33}), \tag{11b}$$

$$\tilde{c}_{1333} = \frac{\sin(2\alpha)}{2}\left(\cos^2(\alpha)\Delta c_{33} - \sin^2(\alpha)\Delta c_{11}\right), \tag{11c}$$

$$\tilde{c}_{2233} = \sin^2(\alpha) c_{1122} + \cos^2(\alpha) c_{1133}, \tag{11d}$$

$$\tilde{c}_{3333} = \cos^2(\alpha) c_{3333} + \sin^2(\alpha) c_{1111} - \frac{\sin^2(2\alpha)}{4}(\Delta c_{11} + \Delta c_{33}). \tag{11e}$$

In Eqs.(10) we introduced the anisotropy factors of the flexoelectric and elastic stiffness tensors

$$\Delta f_{11} = f_{1111} - f_{3311} - f_{1313} - f_{1331}, \quad \Delta f_{33} = f_{3333} - f_{1133} - f_{1313} - f_{1331}., \tag{12a}$$

$$\Delta c_{11} = c_{1111} - c_{1133} - 2c_{1313}, \quad \Delta c_{33} = c_{3333} - c_{1133} - 2c_{1313}.. \tag{12b}$$

For the case of Eq.(2b) application, $f_{1313} = f_{3311} = f_{3113} = f_{2323} = f_{3322} = f_{3223}$, and so expressions (12a) simplify as $\Delta f_{11} = f_{1111} - 2f_{3311} - f_{1313}$, $\Delta f_{33} = f_{3333} - 2f_{1133} - f_{1313}$.

For a **tetragonal crystal,** (symmetry group **4'2m, 422, 4mm, 4/mmm**) the angular dependence of the effective constants $f_{13}^{eff}$ and $f_{23}^{eff}$ is shown in **Fig. 3 and Fig. 4** with the examples of $BaTiO_3$ and $Ba_{0.5}Sr_{0.5}TiO_3$ plates respectively.

As above in Fig.2a, angular dependence $f_{ij}^{eff}(\alpha)$ in **Fig.3(a)** and **Fig.4(a),** follows Eqs.(6), (10)-(12) and (5a), without hidden index-permutation symmetry condition (2b) involved. Used in our calculations coefficients $f_{ijkl}$ and $c_{ijkl}$ are from **Table IV.** According to Ponomareva et.al. [27], for $Ba_{0.5}Sr_{0.5}TiO_3$ we have $f_{1111} \sim f_{1122} \sim 4$ V, while for $BaTiO_3$, according to Maranganti and Sharma [28], $f_{1111} << |f_{1122}| \sim 4$ V. Moreover, in these materials signs of $f_{ijkl}$ are different (compare the first and the second lines in **Table IV**). Therefore, for plates of $BaTiO_3$ and $Ba_{0.5}Sr_{0.5}TiO_3$ we expect different effective flexoelectric responses $f_{ij}^{eff}(\alpha)$. From the table, obtained from *ab initio* calculations at 0 K in cubic approximation [27, 28], the difference $f_{1122} - f_{1212}$ is nonzero and rather high, of the order of (1 ÷ 3) V. Notably, performed for a finite cell, DFT calculations, even under periodic boundary conditions, do not reproduce the hidden index-permutation symmetry of the flexoelectric tensor.

Shown in **Fig.3(b)** and **Fig.4(b),** angular dependences $f_{ij}^{eff}(\alpha)$, were calculated from Eqs.(6), (10)-(12) and (5b), for an infinite crystal where the condition (2b) of hidden index-permutation symmetry applies. For **Fig.3(b)** we used symmetrized off-diagonal coefficients, $f_{1122} = f_{1212} = 1.585$ V, and the diagonal value $f_{1111} = 5.12$ V calculated for $Ba_{0.5}Sr_{0.5}TiO_3$ by Ponomareva et.al. [27]**.** For **Fig.4(b)** the off-diagonal coefficients, $f_{1122} = f_{1212} = -0.935$ V, and the diagonal one is $f_{1111} = 0.04$ V, calculated for $BaTiO_3$ by Maranganti and Sharma [28].

Comparing **Fig.3(a)** with **3(b)** we can conclude that both $f_{13}^{eff}$ and $f_{23}^{eff}$ are strongly affected by the presence of $f_{ijkl}$ hidden symmetry, at that the changes of $f_{ij}^{eff}(\alpha)$ are very different for solid and dashed curves in the figures. Namely, the presence of the $f_{ijkl}$ hidden symmetry changes the angular dependence $f_{ij}^{eff}(\alpha)$, e.g. decreases the number of lobes and decrease the differences between $f_{13}^{eff}$ and $f_{23}^{eff}$.

Comparing **Fig.4(a)** with **4(b)**, we see that both flexoelectric components, $f_{13}^{eff}$ and $f_{23}^{eff}$, are affected by the presence of the index-permutation symmetry (2b) of $f_{ijkl}$, at that the changes of $f_{ij}^{eff}(\alpha)$ are very different for solid ( $f_{13}^{eff}$ ) and dashed ( $f_{23}^{eff}$ ) curves in the figures. The presence of the hidden index-permutation symmetry of $f_{ijkl}$ makes the $\alpha$-dependences of $f_{13}^{eff}$ and $f_{23}^{eff}$ more complex; and their small additional lobes are growing up for both components of the flexoresponse [compare solid and dashed curves in the figure]. However the differences between solid and dashed curves in **Fig.4** [as well as between **Fig.4(a)** and **4(b)**], are much smaller than the differences between solid and dashed curves in **Fig.3**.

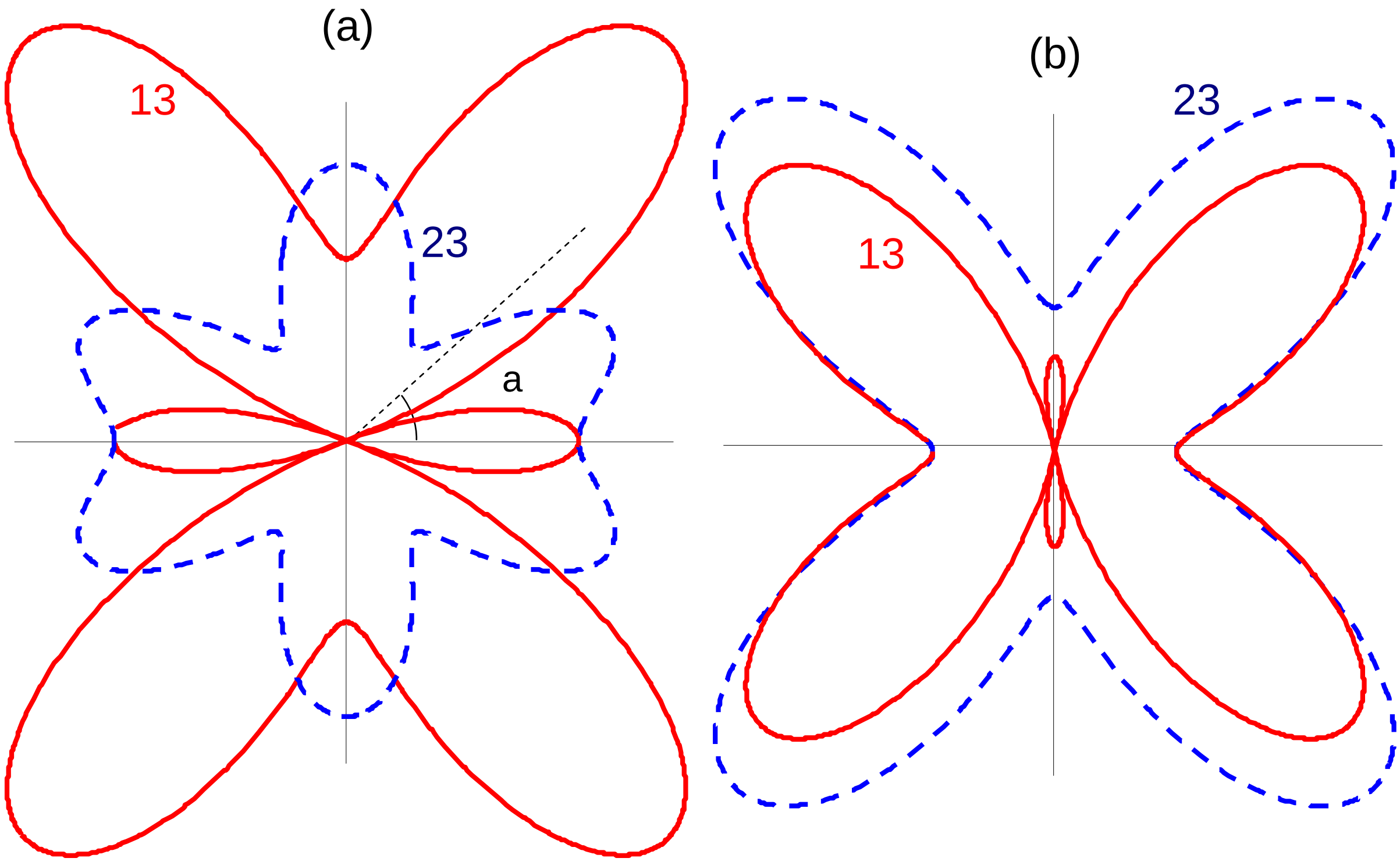

**FIGURE 3.** Angular dependence of the effective flexoresponse components for the $Ba_{0.5}Sr_{0.5}TiO_3$ plate **(a)** and infinite crystal **(b)** of 4mm symmetry. The angle $\alpha$ represents rotation of the coordinate frame with respect to the plate-related crystallographic pseudo-cubic axis $\tilde{X}_2 \circ X_2$ (see **Fig. 1**). Solid and dashed curves represent the flexoresponse in two perpendicular directions, $f_{13}^{eff}$ and $f_{23}^{eff}$, respectively. Values of $f_{ijkl}$ and $c_{ijkl}$ are given in the **Table IV.** The plot **(a)** is calculated from Eqs.(5a). For the plot **(b)**, calculated from Eqs.(5b), we use symmetrized off-diagonal flexoelectric coefficients $f_{1122}=f_{1212}=$ 1.585 V.

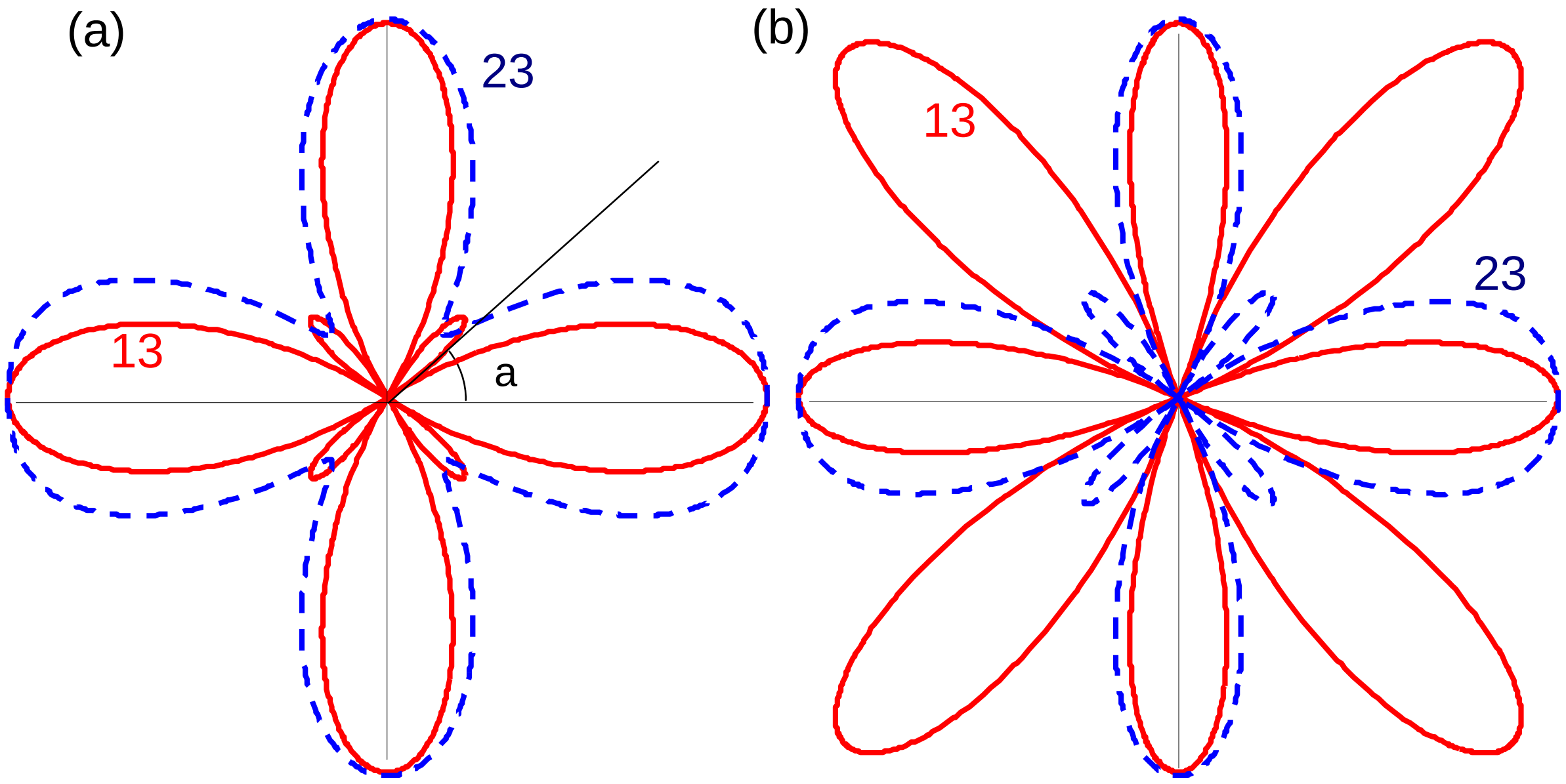

**FIGURE 4.** Angular dependence of the effective flexoresponse components for the $BaTiO_3$ plate **(a)** and infinite crystal **(b)** of 4mm symmetry. The angle $\alpha$ represents rotation of the coordinate frame with respect to the plate-related crystallographic pseudo-cubic axis $\tilde{X}_2$ o $X_2$ (see **Fig. 1**). Solid and dashed curves represent the flexoresponse in two perpendicular directions, $f_{13}^{eff}$ and $f_{23}^{eff}$, respectively. Values of $f_{ijkl}$ and $c_{ijkl}$ are given in the **Table IV.** The plot **(a)** is calculated from Eqs.(5a). For the plot **(b)**, calculated from Eqs.(5b), we use symmetrized off-diagonal flexoelectric coefficients $f_{1122}$=$f_{1212}$= 1.585 V.

**Table IV.** Flexoelectric ($f_{ijkl}$) and elastic stiffness ($c_{ijkl}$) tensor components of $Ba_xSr_{(1-x)}TiO_3$ ($x$ = 0.5, 1)

| tensor | values of components | | | | | | Reference & notation |
|---|---|---|---|---|---|---|---|
| $f_{ijkl}$ (V) | $f_{1111}$=5.12 | $f_{1122}$=3.32 | n/a | n/a | $f_{1212}$=0.05 | n/a | Ponomareva et.al. [27], ab initio calculations at 0 K in cubic approximation $Ba_{0.5}Sr_{0.5}TiO_3$ |
| $f_{ijkl}$ (V) | $f_{1111}$=0.04 | $f_{1122}$=- 1.39 | n/a | n/a | $f_{1212}$=- 0.48 | n/a | Maranganti and Sharma [28] ab calculations at 0 K in cubic approximation for $BaTiO_3$ |
| $c_{ijkl}$ ($10^{11}$ Pa) | $c_{1111}$=2.43 | $c_{1122}$=1.28 | $c_{1133}$=1.23 | $c_{1212}$=1.20 | $c_{1313}$=0.55 | $c_{3333}$=1.48 | Shaefer et al [30], experiment at RT for $BaTiO_3$ |

RT - room temperature

The effective flexoresponse of $BaTiO_3$ in **orthorhombic phase** will be considered elsewhere, due to the unknown flexoelectric coefficients in this case.

To resume the section, the hidden index-permutation symmetry of $f_{ijkl}$ strongly affect on the angular dependences of the effective flexoresponse $f_{ij}^{eff}(\alpha)$. The result can be of direct use for the unambiguous determination of $f_{ijkl}$ from bending experiments. However in appeared that the differences between the angular dependences calculated with and without index-permutation symmetry (2b) significantly depend on the numerical values of $f_{ijkl}$. However the condition (2b), $f_{ijkl} = f_{ilkj}$, valid for infinite material, does not hold neither in experiments [25, 26] conducted for finite thin plates, nor in *ab initio* calculations [27, 28] performed for a 3D-confined computation cell. Possible ways how to verify our prediction are discussed in the next section.

## IV. IMPACT OF THE HIDDEN SYMMETRY ON THE DETERMINATION OF THE FLEXOCOUPLING CONSTANTS

We should ascertain that the available experimental data [25, 26], *ab initio* calculations [27, 28] are in contradiction with the symmetry predictions [11]. There is a seeming contradiction between the measured and calculated values of the flexocoupling tensor that was firstly notified by Biancoli et al [31]. Moreover, the upper limits for the values $f_{ijkl}$ established by Yudin et al [32], as well as the values calculated from the first principles for bulk ferroics [33, 34, 35, 36, 37], can be several orders of magnitude smaller than those measured experimentally in ferroelectric ceramics [38, 39, 40] and thin films [41], ferroelectric relaxor polymers [42] and electrets [43], incipient ferroelectrics [26] and biological membranes [44, 45]. Stengel [46], Abdollahi et al. [47] and Rahmati et al. [48] argued that the giant values of the flexoelectric effect, measured sometimes, can originate from the different electric boundary conditions. Bersuker argued [49] that the anomalously high flexoelectric coefficients in perovskite ceramics may be related with the manifestation of the pseudo Jahn-Teller effect. Steaming from a vibronic nature [50], pseudo Jahn-Teller effect can affect the dynamic flexoeffect in ferroics. Indeed, the available information about the numerical values of $M_{ij}$ is completely controversial. On one hand, there are microscopic theories in which the dynamic effect is absent [46], and, on the other hand, its determination from the soft phonon spectra leads to nonzero $M_{ij}$ [51, 52], and the impact of dynamic flexoelectric effect appeared comparable to that of the static one.

Notably that experimental measurements of the soft phonon dispersion by inelastic neutron and Raman scattering, and theoretical calculations allow one to extract the valuable information about the flexoelectric coefficients [53, 54, 55]. At the same time the scattering spectra can be readily obtained for the plate or beam, and for a bulk material, in order to exclude any size effect. To illustrate the statement let us analyze analytical expressions derived in Ref.[51] for a bulk material. The Fourier $\mathbf{k}$-$\omega$ spectra of inverse dynamic linear susceptibility $\tilde{\chi}_{ij}(\mathbf{k},\omega)$ relates elastic displacement and polarization fluctuations with external electric field and elastic force spectra in the following way:

$$\begin{pmatrix}\delta\tilde{U}_i(\mathbf{k},\omega)\\ \delta\tilde{P}_i(\mathbf{k},\omega)\end{pmatrix}=\tilde{\chi}_{ij}(\mathbf{k},\omega)\begin{pmatrix}\delta\tilde{F}_j^{ext}(\mathbf{k},\omega)\\ \delta\tilde{E}_j^{ext}(\mathbf{k},\omega)\end{pmatrix}, \qquad (13)$$

where the indexes $i, j = 1, 2, 3$. The expression for the matrix $6\times 6$ of inverse susceptibility, $\tilde{\chi}_{ij}^{-1}(\mathbf{k},\omega)$, valid for a dielectric or paraelectric solid, can be presented in a block form:

$$\tilde{\chi}_{ij}^{-1}(\mathbf{k},\omega)=\begin{pmatrix} c_{imjl}k_l k_m-\rho\omega^2\delta_{ij} & f_{iljk}k_l k_m - M_{ij}\omega^2 \\ f_{jnli}k_l k_n - M_{ij}\omega^2 & \left(2\alpha-\mu\omega^2\right)\delta_{ij}+\dfrac{k_i k_j}{\varepsilon_b\varepsilon_0 k^2}+g_{imjl}k_m k_l \end{pmatrix} \qquad (14)$$

Each of the four elements in Eq.(14) is a matrix $3\times 3$, and therefore the comprehensive analytical form of direct matrix $\tilde{\chi}_{ij}(\mathbf{k},\omega)$ cannot be derived in a general case of arbitrary point symmetry and arbitrary $\mathbf{k}$-orientation. The analytical expressions for $\tilde{\chi}_{ij}(\mathbf{k},\omega)$ are available only in several particular cases [51].

Note that the experimental determination of $\tilde{\chi}_{ij}(\mathbf{k},\omega)$ components can open the way for the indirect determination of poorly known static and dynamic flexocoupling tensors, $f_{ijkl}$ and $M_{ij}$ (see e.g. [51-55]). As an example, for a cubic point symmetry three constants of the static flexoeffect ( $f_{1111}$, $f_{1212}$ and $f_{1122}$ ) and one constant of dynamic flexoeffect ( $M_{11}=M_{22}=M_{33}$ ) can be determined from the fitting of three acoustic and one optic soft phonon modes at a given $\mathbf{k}$-direction. If the hidden symmetry (2b) exists, the best fitting corresponds to the case $f_{1212}=f_{1122}$ independently on $\mathbf{k}$-direction.

Hence it is very important to know the quantitative contribution of the flexocoupling to $\tilde{\chi}(\mathbf{k},\omega)$. Thus the role of the hidden index-permutation symmetry (2b) of $f_{ijkl}$ can be unique important, because its application can significantly simplify the fitting of the spectra due to the minimization of $f_{ijkl}$ independent components, and facilitating the way for its unambiguous determination from the scattering experiments.

## V. SUMMARY

- Using direct matrix method [see Eqs.(5)] we established the structure, including the number of nonzero and independent elements, of the static flexocoupling tensor $f_{ijkl}$ for all 32 point groups using point symmetry, previously known evident symmetry [ $f_{ijkl} \equiv f_{jikl}$, see Eq.(2a)] and recently established hidden index-permutation symmetry [ $f_{ijkl} = f_{ilkj}$ , see Eq.(2b)]. We performed the comparison of these results and demonstrated that the hidden index-permutation symmetry reduces significantly the number of nonzero independent elements of $f_{ijkl}$.
- We list the explicit form of the tensor $f_{ijkl}$ [see Tables I and II] and the flexocoupling energy in the form of Lifshitz invariant, $\frac{f_{ijkl}}{2}\left(P_k\frac{\partial u_{ij}}{\partial x_l} - u_{ij}\frac{\partial P_k}{\partial x_l}\right)$ [see Appendix A2] for several cubic, tetragonal and orthorhombic point symmetries using both direct method and group-theoretical approach, which lead to the same results. For these symmetries we visualize the effective flexoresponse of a thin plate allowing for point symmetry, evident and hidden index-permutation symmetries, analyze its anisotropy and angular dependences. These results can be of direct use for the unambiguous determination of $f_{ijkl}$ from bending experiments.
- We discuss the problem of the flexocoupling constants determination from inelastic neutron and Raman scattering experiments. We predicted that the hidden index-permutation symmetry (2b) can significantly simplify the fitting procedure due to the minimization of $f_{ijkl}$ independent components, and so it opens the way for $f_{ijkl}$ unambiguous determination from the scattering experiments.

## DESCRIPTION OF SUPPLEMENTARY MATERIALS

**APPENDIX A1 –** flexoelectric tensor for all 32 point symmetry groups with and without the "hidden" index-permutation symmetry, and their comparison

**APPENDIX A2 –**application of the flexoelectric tensor symmetry to the Lifshitz invariant (1) for several point symmetries

**APPENDIX B –** group-theoretical approach to the explicit form of flexocoupling energy, for the symmetry groups *m*3*m*, 4/*mmm* and *mmm*

**APPENDIX C –** effective flexoelectric response of a thin plate (3 knifes experiment) on example of cubic and tetragonal symmetry

**Authors' contribution.** E.A.E. jointly with V.V.K. performed calculations of the flexoelectric tensor symmetry using direct matrix method (Appendix A1). Also E.A.E. performed all analytical calculations of effective flexoresponse in section III and Appendix C, and generated figures 1-3. V.P. performed calculations of the flexoelectric tensor symmetry and flexocoupling energy using group-theoretical approach (Appendix B). A.N.M. generated research idea, performed calculations of flexocoupling energy using direct method (Appendix A2), and calculations in section IV, analyze obtained results and wrote the manuscript. All authors contributed equally to the results discussion and manuscript improvement.

[14] E.A. Eliseev, M.D. Glinchuk, V. Khist, V.V. Skorokhod, R. Blinc, A.N. Morozovska. Linear magnetoelectric coupling and ferroelectricity induced by the flexomagnetic effect in ferroics. Phys.Rev. **B 84**, 174112 (2011)

[15] Supplementary Materials

[16] W. Cochran. Dynamical, scattering and dielectric properties of ferroelectric crystals, Advances in Physics, 18:72, 157-192 (1969).

[17] J. Hlinka, I. Gregora, and V. Vorlıcek. Complete spectrum of long-wavelength phonon modes in Sn2P2S6 by Raman scattering. Phys.Rev. B **65**, 064308 (2002)

[18] R.M.Yevych, Yu. M. Vysochanskii, M.M. Khoma and S.I. Perechinskii. Lattice instability at phase transitions near the Lifshitz point in proper monoclinic ferroelectricsJ. Phys.: Condens. Matter **18**, 4047–4064 (2006)

[19] G. Shirane, J. D. Axe, J. Harada, and J. P. Remeika. "Soft ferroelectric modes in lead titanate." *Physical Review B* 2, no. 1: 155 (1970)

[20] G. Shirane, J. D. Axe, J. Harada, and A. Linz. "Inelastic Neutron Scattering from Single-Domain $BaTiO_3$. *Physical Review B* 2, no. 9: 3651 (1970).

[21] G. Shirane, B. C. Frazer, V. J. Minkiewicz, J. A. Leake, and A. Linz. "Soft Optic Modes in Barium Titanate" Phys. Rev. Lett. 19, 234 (1967)

[22] S. W. H.,Eijt, R. Currat, J. E. Lorenzo, P. Saint-Gregoire, B. Hennion, and Yu M. Vysochanskii. Soft modes and phonon interactions in Sn2P2S6 studied by neutron scattering. The European Physical Journal B-Condensed Matter and Complex Systems **5**, no. 2: 169-178 (1998).

[23] S. W. H., Eijt, R. Currat, J. E. Lorenzo, P. Saint-Gregoire, S. Katano, T. Janssen, B. Hennion, and Yu M. Vysochanskii. Soft modes and phonon interactions in $Sn_2P_2Se_6$ studied by means of neutron scattering. Journal of Physics: Condensed Matter **10**, no. 22: 4811 (1998).

[24] E.A. Eliseev, A.N. Morozovska, M.D. Glinchuk, B.Y. Zaulychny, V.V. Skorokhod, R. Blinc. Surface-induced piezomagnetic, piezoelectric, and linear magnetoelectric effects in nanosystems. Phys. Rev. **B. 82**, 085408 (2010).

[25] P. Zubko, G. Catalan, A. Buckley, P.R. L. Welche, J. F. Scott. Strain-gradient-induced polarization in $SrTiO_3$ single crystals, Phys. Rev. Lett. **99**, 167601 (2007).

[26] P. Zubko, G. Catalan, A. Buckley, P. R. L. Welche, and J. F. Scott, Erratum: Strain-Gradient-Induced Polarization in SrTiO3 Single Crystals [Phys. Rev. Lett. 99, 167601 (2007)], Phys. Rev. Lett. 100, 199906 (2008).

[27] I. Ponomareva, A.K. Tagantsev, L. Bellaiche. Phys. Rev. B **85**, Finite-temperature flexoelectricity in ferroelectric thin films from first principles, 104101 (2012).

[28] R. Maranganti and P. Sharma. Phys. Rev. B **80**, 054109 (2009).

[29] R. O. Bell, and Georg Rupprecht. "Elastic constants of strontium titanate." *Physical Review* 129, no. 1: 90 (1963).

[48]Amir Hossein Rahmati, Shengyou Yang, Siegfried Bauer and Pradeep Sharma. Nonlinear bending deformation of soft electrets and prospects for engineering flexoelectricity and transverse (d31) piezoelectricity, Soft Matter, **15**, 127 (2019)

[49] I. B. Bersuker. Pseudo Jahn–Teller effect in the origin of enhanced flexoelectricity. Appl. Phys. Lett. 106, 022903 (2015); doi: 10.1063/1.4905679

[50] V. Polinger, and I. B. Bersuker. Pseudo Jahn-Teller effect in permittivity of ferroelectric Perovskites. IOP Conf. Series: Journal of Physics: Conf. Series 833, 012012 (2017)

[51] Anna N. Morozovska, Yulian M. Vysochanskii, Oleksandr V. Varenyk, Maxim V. Silibin, Sergei V. Kalinin, and Eugene A. Eliseev. Flexocoupling impact on the generalized susceptibility and soft phonon modes in the ordered phase of ferroics, Phys. Rev. B 92, 094308 (2015).

[52] Anna N. Morozovska, Eugene A. Eliseev, Christian M. Scherbakov, and Yulian M. Vysochanskii, The influence of elastic strain gradient on the upper limit of flexocoupling strength, spatially-modulated phases and soft phonon dispersion in ferroics, Phys. Rev. B 94, 174112 (2016).

[53] Anna N. Morozovska, Maya D. Glinchuk, Eugene A. Eliseev, and Yulian M. Vysochanskii. Flexocoupling-induced soft acoustic mode and the spatially modulated phases in ferroelectrics, Phys. Rev. B, 96, 094111 (2017).

[54] Eugene A. Eliseev, Anna N. Morozovska, Maya D. Glinchuk, and Sergei V. Kalinin, Missed surface waves in non-piezoelectric solids, Phys. Rev. B 96, 045411 (2017).

[55] A. N. Morozovska, and E. A. Eliseev. Size effect of soft phonon dispersion law in nanosized ferroics: "true" size effect of phonon frequency via "apparent" size effect of flexoelectric coupling. Accepted to Phys.Rev. B (2019) (http://arxiv.org/abs/1812.04702)

# SUPPLEMENTARY MATERIALS

## Appendix A1. Structure of flexoelectric tensor allowing for point symmetry, "evident" index-permutation symmetry and "hidden" index-permutation symmetry

Let us consider two 4-th rank tensors of different permutation (internal) symmetry:

### 1. Structure of flexoelectric tensor (FET) allowing for point symmetry (PS) and "evident" index-permutation symmetry (EIPS)

The tensor with "evident" index-permutation symmetry (EIPS) should be invariant to permutation of the first and the second indices (due to the symmetry of strain tensor), so that general relations for EIPS FET are listed below:

$$\bar{f}_{1211} = \bar{f}_{2111}, \bar{f}_{1212} = \bar{f}_{2112}, \bar{f}_{1213} = \bar{f}_{2113}, \quad \text{(A.1a)}$$

$$\bar{f}_{1221} = \bar{f}_{2121}, \bar{f}_{1222} = \bar{f}_{2122}, \bar{f}_{1223} = \bar{f}_{2123}, \quad \text{(A.1b)}$$

$$\bar{f}_{1231} = \bar{f}_{2131}, \bar{f}_{1232} = \bar{f}_{2132}, \bar{f}_{1233} = \bar{f}_{2133}, \quad \text{(A.1c)}$$

$$\bar{f}_{1311} = \bar{f}_{3111}, \bar{f}_{1312} = \bar{f}_{3112}, \bar{f}_{1313} = \bar{f}_{3113}, \quad \text{(A.1d)}$$

$$\bar{f}_{1321} = \bar{f}_{3121}, \bar{f}_{1322} = \bar{f}_{3122}, \bar{f}_{1323} = \bar{f}_{3123}, \quad \text{(A.1e)}$$

$$\bar{f}_{1331} = \bar{f}_{3131}, \bar{f}_{1332} = \bar{f}_{3132}, \bar{f}_{1333} = \bar{f}_{3133}, \quad \text{(A.1f)}$$

$$\bar{f}_{2311} = \bar{f}_{3211}, \bar{f}_{2312} = \bar{f}_{3212}, \bar{f}_{2313} = \bar{f}_{3213} \quad \text{(A.1g)}$$

$$\bar{f}_{2321} = \bar{f}_{3221}, \bar{f}_{2322} = \bar{f}_{3222}, \bar{f}_{2323} = \bar{f}_{3223}, \quad \text{(A.1h)}$$

$$\bar{f}_{2331} = \bar{f}_{3231}, \bar{f}_{2332} = \bar{f}_{3232}, \bar{f}_{2333} = \bar{f}_{3233}. \quad \text{(A.1i)}$$

Point symmetry of the media imposes additional restrictions of the form of tensor [see Eqs.(5a) in the main text and Table AI below].

**Table AI.** Number of elements of EIPS FET for different point groups

| point group | non-zero | independent | nonzero element and relations between them |
|---|---|---|---|
| m3m, 432 $\bar{4}3m$ | 21 | 3 | $\bar{f}_{1111} = \bar{f}_{2222} = \bar{f}_{3333}$; $\bar{f}_{1122} = \bar{f}_{2211} = \bar{f}_{1133} = \bar{f}_{3311} = \bar{f}_{2233} = \bar{f}_{3322}$; $\bar{f}_{1212} = \bar{f}_{2112} = \bar{f}_{1221} = \bar{f}_{2121} =$ $\bar{f}_{1313} = \bar{f}_{3113} = \bar{f}_{1331} = \bar{f}_{3131}$ $\bar{f}_{2323} = \bar{f}_{3223} = \bar{f}_{2332} = \bar{f}_{3232}$; |
| 23, m3, | 21 | 5 | $\bar{f}_{1111} = \bar{f}_{2222} = \bar{f}_{3333}$; $\bar{f}_{1122} = \bar{f}_{2233} = \bar{f}_{3311}$; $\bar{f}_{1212} = \bar{f}_{2112} = \bar{f}_{1331} = \bar{f}_{3131} = \bar{f}_{2323} = \bar{f}_{3223}$; $\bar{f}_{1133} = \bar{f}_{2211} = \bar{f}_{3322}$; $\bar{f}_{1221} = \bar{f}_{2121} = \bar{f}_{1313} = \bar{f}_{3113} = \bar{f}_{2332} = \bar{f}_{3232}$ |
| $\bar{6}m2$,<br>622,<br>6mm,<br>6/mmm | 21 | 7 | $\bar{f}_{1111} = \bar{f}_{2222}$; $\bar{f}_{1122} = \bar{f}_{2211}$; $\bar{f}_{1133} = \bar{f}_{2233}$;<br>$\bar{f}_{1212} = \bar{f}_{2112} = \bar{f}_{1221} = \bar{f}_{2121} = \frac{\bar{f}_{1111}}{2} - \frac{\bar{f}_{1122}}{2}$<br>$\bar{f}_{1313} = \bar{f}_{3113} = \bar{f}_{2323} = \bar{f}_{2323}$;<br>$\bar{f}_{1331} = \bar{f}_{3131} = \bar{f}_{2332} = \bar{f}_{3232}$;<br>$\bar{f}_{3311} = \bar{f}_{3322}$; $\bar{f}_{3333}$; |
| 6, $\bar{6}$,<br>6/m | 39 | 12 | $\bar{f}_{1111} = \bar{f}_{2222}$; $\bar{f}_{1112} = -\bar{f}_{2221}$;<br>$\bar{f}_{1121} = -\bar{f}_{2212}$; $\bar{f}_{1122} = \bar{f}_{2211}$; $\bar{f}_{1133} = \bar{f}_{2233}$;<br>$\bar{f}_{1211} = \bar{f}_{2111} = -\bar{f}_{1222} = -\bar{f}_{2122} = -\frac{\bar{f}_{1112}}{2} - \frac{\bar{f}_{1121}}{2}$;<br>$\bar{f}_{1212} = \bar{f}_{2112} = \bar{f}_{1221} = \bar{f}_{2121} = \frac{\bar{f}_{1111}}{2} - \frac{\bar{f}_{1122}}{2}$;<br>$\bar{f}_{1313} = \bar{f}_{3113} = \bar{f}_{2323} = \bar{f}_{2323}$;<br>$\bar{f}_{1323} = \bar{f}_{3123} = -\bar{f}_{2313} = -\bar{f}_{3213}$;<br>$\bar{f}_{1331} = \bar{f}_{3131} = \bar{f}_{2332} = \bar{f}_{3232}$;<br>$\bar{f}_{1332} = \bar{f}_{3132} = -\bar{f}_{2331} = -\bar{f}_{3231}$;<br>$\bar{f}_{3312} = -\bar{f}_{3321}$; $\bar{f}_{3311} = \bar{f}_{3322}$; $\bar{f}_{3333}$; |
| 3, $\bar{3}$ | 71 | 18 | $\bar{f}_{1111} = \bar{f}_{2222}$; $\bar{f}_{1112} = -\bar{f}_{2221}$; $\bar{f}_{1113} = -\bar{f}_{1223} = -\bar{f}_{2123} = -\bar{f}_{2213}$;<br>$\bar{f}_{1121} = -\bar{f}_{2212}$; $\bar{f}_{1122} = \bar{f}_{2211}$; $\bar{f}_{1123} = \bar{f}_{1213} = \bar{f}_{2113} = -\bar{f}_{2223}$;<br>$\bar{f}_{1131} = -\bar{f}_{1232} = -\bar{f}_{2132} = -\bar{f}_{2231}$;<br>$\bar{f}_{1132} = \bar{f}_{1231} = \bar{f}_{2131} = -\bar{f}_{2232}$; $\bar{f}_{1133} = \bar{f}_{2233}$;<br>$\bar{f}_{1211} = \bar{f}_{2111} = -\bar{f}_{1222} = -\bar{f}_{2122} = -\frac{\bar{f}_{1112}}{2} - \frac{\bar{f}_{1121}}{2}$;<br>$\bar{f}_{1212} = \bar{f}_{2112} = \bar{f}_{1221} = \bar{f}_{2121} = \frac{\bar{f}_{1111}}{2} - \frac{\bar{f}_{1122}}{2}$;<br>$\bar{f}_{1311} = \bar{f}_{3111} = -\bar{f}_{1322} = -\bar{f}_{3122} = -\bar{f}_{2321} = -\bar{f}_{3221} = -\bar{f}_{2312} = -\bar{f}_{3212}$;<br>$\bar{f}_{1312} = \bar{f}_{1321} = \bar{f}_{2311} = \bar{f}_{3112} = \bar{f}_{3121} = \bar{f}_{3211} = -\bar{f}_{2322} = -\bar{f}_{3222}$;<br>$\bar{f}_{1313} = \bar{f}_{3113} = \bar{f}_{2323} = \bar{f}_{2323}$;<br>$\bar{f}_{1323} = \bar{f}_{3123} = -\bar{f}_{2313} = -\bar{f}_{3213}$;<br>$\bar{f}_{1331} = \bar{f}_{3131} = \bar{f}_{2332} = \bar{f}_{3232}$;<br>$\bar{f}_{1332} = \bar{f}_{3132} = -\bar{f}_{2331} = -\bar{f}_{3231}$;<br>$\bar{f}_{3312} = -\bar{f}_{3321}$; $\bar{f}_{3311} = \bar{f}_{3322}$; $\bar{f}_{3333}$; |
| 32, 3m,<br>$\bar{3}m$ | 37 | 10 | $\bar{f}_{1111} = \bar{f}_{2222}$; $\bar{f}_{1113} = -\bar{f}_{1223} = -\bar{f}_{2123} = -\bar{f}_{2213}$;<br>$\bar{f}_{1122} = \bar{f}_{2211}$; $\bar{f}_{1131} = -\bar{f}_{1232} = -\bar{f}_{2132} = -\bar{f}_{2231}$; $\bar{f}_{1133} = \bar{f}_{2233}$;<br>$\bar{f}_{1212} = \bar{f}_{2112} = \bar{f}_{1221} = \bar{f}_{2121} = \frac{\bar{f}_{1111}}{2} - \frac{\bar{f}_{1122}}{2}$;<br>$\bar{f}_{1311} = \bar{f}_{3111} = -\bar{f}_{1322} = -\bar{f}_{3122} = -\bar{f}_{2321} = -\bar{f}_{3221} = -\bar{f}_{2312} = -\bar{f}_{3212}$;<br>$\bar{f}_{1313} = \bar{f}_{3113} = \bar{f}_{2323} = \bar{f}_{2323}$;<br>$\bar{f}_{1331} = \bar{f}_{3131} = \bar{f}_{2332} = \bar{f}_{3232}$;<br>$\bar{f}_{3311} = \bar{f}_{3322}$; $\bar{f}_{3333}$; |
| $\bar{4}\,2m$,<br>422,<br>4mm,<br>4/mmm | 21 | 8 | $\bar{f}_{1111} = \bar{f}_{2222}$; $\bar{f}_{3333}$;<br>$\bar{f}_{1122} = \bar{f}_{2211}$; $\bar{f}_{1133} = \bar{f}_{2233}$; $\bar{f}_{3311} = \bar{f}_{3322}$;<br>$\bar{f}_{1212} = \bar{f}_{2112} = \bar{f}_{1221} = \bar{f}_{2121}$;<br>$\bar{f}_{1313} = \bar{f}_{3113} = \bar{f}_{2323} = \bar{f}_{3223}$;<br>$\bar{f}_{1331} = \bar{f}_{3131} = \bar{f}_{2332} = \bar{f}_{3232}$; |
| 4, $\bar{4}$, 4/m | 39 | 14 | $\bar{f}_{1111} = \bar{f}_{2222}$; $\bar{f}_{1122} = \bar{f}_{2211}$; $\bar{f}_{1133} = \bar{f}_{2233}$; $\bar{f}_{3311} = \bar{f}_{3322}$;<br>$\bar{f}_{1212} = \bar{f}_{2112} = \bar{f}_{1221} = \bar{f}_{2121}$;<br>$\bar{f}_{1313} = \bar{f}_{3113} = \bar{f}_{2323} = \bar{f}_{3223}$;<br>$\bar{f}_{1331} = \bar{f}_{3131} = \bar{f}_{2332} = \bar{f}_{3232}$;<br>$\bar{f}_{1211} = \bar{f}_{2111} = -\bar{f}_{1222} = -\bar{f}_{2122}$;<br>$\bar{f}_{1323} = \bar{f}_{3123} = -\bar{f}_{2313} = -\bar{f}_{3213}$<br>$\bar{f}_{1332} = \bar{f}_{3132} = -\bar{f}_{2331} = -\bar{f}_{3231}$;<br>$\bar{f}_{1112} = -\bar{f}_{2221}$; $\bar{f}_{1121} = -\bar{f}_{2212}$;<br>$\bar{f}_{3312} = -\bar{f}_{3321}$; $\bar{f}_{3333}$; |
| 222, mm2, mmm | 21 | 15 | $\bar{f}_{1111}$; $\bar{f}_{1122}$; $\bar{f}_{1133}$;<br>$\bar{f}_{2211}$; $\bar{f}_{2222}$; $\bar{f}_{2233}$;<br>$\bar{f}_{3311}$; $\bar{f}_{3322}$; $\bar{f}_{3333}$;<br>$\bar{f}_{1212} = \bar{f}_{2112}$; $\bar{f}_{1221} = \bar{f}_{2121}$;<br>$\bar{f}_{1313} = \bar{f}_{3113}$; $\bar{f}_{1331} = \bar{f}_{3131}$;<br>$\bar{f}_{2323} = \bar{f}_{3223}$; $\bar{f}_{2332} = \bar{f}_{3232}$; |
| 2, m, 2/m | 41 | 28 | $\bar{f}_{1111}$; $\bar{f}_{1112}$; $\bar{f}_{1121}$; $\bar{f}_{1122}$; $\bar{f}_{1133}$;<br>$\bar{f}_{2211}$; $\bar{f}_{2212}$; $\bar{f}_{2221}$; $\bar{f}_{2222}$; $\bar{f}_{2233}$;<br>$\bar{f}_{3311}$; $\bar{f}_{3312}$; $\bar{f}_{3321}$; $\bar{f}_{3322}$; $\bar{f}_{3333}$;<br>$\bar{f}_{1212} = \bar{f}_{2112}$; $\bar{f}_{1221} = \bar{f}_{2121}$; $\bar{f}_{1211} = \bar{f}_{2111}$; $\bar{f}_{1222} = \bar{f}_{2122}$; $\bar{f}_{1233} = \bar{f}_{2133}$;<br>$\bar{f}_{1313} = \bar{f}_{3113}$; $\bar{f}_{1331} = \bar{f}_{3131}$; $\bar{f}_{1323} = \bar{f}_{3123}$; $\bar{f}_{1332} = \bar{f}_{3132}$;<br>$\bar{f}_{2313} = \bar{f}_{3213}$; $\bar{f}_{2323} = \bar{f}_{3223}$; $\bar{f}_{2331} = \bar{f}_{3231}$; $\bar{f}_{2332} = \bar{f}_{3232}$; |
| 1, $\bar{1}$ | 81 | 54 | $\bar{f}_{1111}, \bar{f}_{1112}, \bar{f}_{1113}, \bar{f}_{1121}, \bar{f}_{1122}, \bar{f}_{1123}, \bar{f}_{1131}, \bar{f}_{1132}, \bar{f}_{1133}$;<br>$\bar{f}_{1211} = \bar{f}_{2111}, \bar{f}_{1212} = \bar{f}_{2112}, \bar{f}_{1213} = \bar{f}_{2113}$,<br>$\bar{f}_{1221} = \bar{f}_{2121}, \bar{f}_{1222} = \bar{f}_{2122}, \bar{f}_{1223} = \bar{f}_{2123}$,<br>$\bar{f}_{1231} = \bar{f}_{2131}, \bar{f}_{1232} = \bar{f}_{2132}, \bar{f}_{1233} = \bar{f}_{2133}$,<br>$\bar{f}_{1311} = \bar{f}_{3111}, \bar{f}_{1312} = \bar{f}_{3112}, \bar{f}_{1313} = \bar{f}_{3113}$,<br>$\bar{f}_{1321} = \bar{f}_{3121}, \bar{f}_{1322} = \bar{f}_{3122}, \bar{f}_{1323} = \bar{f}_{3123}$,<br>$\bar{f}_{1331} = \bar{f}_{3131}, \bar{f}_{1332} = \bar{f}_{3132}, \bar{f}_{1333} = \bar{f}_{3133}$,<br>$\bar{f}_{2211}, \bar{f}_{2212}, \bar{f}_{2213}, \bar{f}_{2221}, \bar{f}_{2222}, \bar{f}_{2223}, \bar{f}_{2231}, \bar{f}_{2232}, \bar{f}_{2233}$;<br>$\bar{f}_{2311} = \bar{f}_{3211}, \bar{f}_{2312} = \bar{f}_{3212}, \bar{f}_{2313} = \bar{f}_{3213}$,<br>$\bar{f}_{2321} = \bar{f}_{3221}, \bar{f}_{2322} = \bar{f}_{3222}, \bar{f}_{2323} = \bar{f}_{3223}$,<br>$\bar{f}_{2331} = \bar{f}_{3231}, \bar{f}_{2332} = \bar{f}_{3232}, \bar{f}_{2333} = \bar{f}_{3233}$,<br>$\bar{f}_{3311}, \bar{f}_{3312}, \bar{f}_{3313}, \bar{f}_{3321}, \bar{f}_{3322}, \bar{f}_{3323}, \bar{f}_{3331}, \bar{f}_{3332}, \bar{f}_{3333}$. |

## 2. Structure of flexoelectric tensor allowing for point symmetry (PS) and "hidden" index-permutation symmetry (HIPS)

The tensor with "hidden" index-permutation symmetry (HIPS) should be invariant to permutation of the first and the second indices (due to the symmetry of strain tensor) and of the second and the fourth indices (due to "new" symmetry of gradients). Therefore, the general relations are two-fold for $f_{ijkl}$ tensor:

1). HIPS FET is invariant to permutation of the first and the second indices:

$$f_{1211} = f_{2111}, f_{1212} = f_{2112}, f_{1213} = f_{2113}, \tag{A.2a}$$

$$f_{1221} = f_{2121}, f_{1222} = f_{2122}, f_{1223} = f_{2123}, \tag{A.2b}$$

$$f_{1231} = f_{2131}, f_{1232} = f_{2132}, f_{1233} = f_{2133}, \tag{A.2c}$$

$$f_{1311} = f_{3111}, f_{1312} = f_{3112}, f_{1313} = f_{3113}, \tag{A.2d}$$

$$f_{1321} = f_{3121}, f_{1322} = f_{3122}, f_{1323} = f_{3123}, \tag{A.2e}$$

$$f_{1331} = f_{3131}, f_{1332} = f_{3132}, f_{1333} = f_{3133}, \tag{A.2f}$$

$$f_{2311} = f_{3211}, f_{2312} = f_{3212}, f_{2313} = f_{3213} \tag{A.2g}$$

$$f_{2321} = f_{3221}, f_{2322} = f_{3222}, f_{2323} = f_{3223}, \tag{A.2h}$$

$$f_{2331} = f_{3231}, f_{2332} = f_{3232}, f_{2333} = f_{3233}, \tag{A.2i}$$

2). HIPS FET is invariant to permutation of the second and the fourth indices:

$$f_{1112} = f_{1211}, f_{1113} = f_{1311}, f_{1122} = f_{1221}, f_{1123} = f_{1321}, f_{1132} = f_{1231}, f_{1133} = f_{1331}, \tag{A.3a}$$

$$f_{1213} = f_{1312}, f_{1223} = f_{1322}, f_{1233} = f_{1332}, \tag{A.3d}$$

$$f_{2112} = f_{2211}, f_{2113} = f_{2311}, f_{2122} = f_{2221}, f_{2123} = f_{2321}, f_{2132} = f_{2231}, f_{2133} = f_{2331}, \tag{A.3b}$$

$$f_{2213} = f_{2312}, f_{2223} = f_{2322}, f_{2233} = f_{2332}, \tag{A.3h}$$

$$f_{3112} = f_{3211}, f_{3113} = f_{3311}, f_{3122} = f_{3221}, , f_{3123} = f_{3321}, \ f_{3132} = f_{3231}, f_{3133} = f_{3331} \tag{A.3c}$$

$$f_{3213} = f_{3312}, f_{3223} = f_{3322}, f_{3233} = f_{3332} \tag{A.3e}$$

Point symmetry of the media imposes additional restrictions of the form of tensor [see Eqs.(5b) in the main text and Table AII below].

**Table AII.** Number of elements of HIPS FET for different point groups

| point group | non-zero | independent | nonzero element and relations between them |
|---|---|---|---|
| m3m, 432, $\bar{4}3m$ | 21 | 2 | $f_{1111} = f_{2222} = f_{3333}$; $f_{1122} = f_{2211} = f_{1133} = f_{3311} = f_{2233} = f_{3322} =$ $f_{1221} = f_{2112} = f_{1331} = f_{3113} = f_{2332} = f_{3223} =$ $f_{1212} = f_{2121} = f_{1313} = f_{3131} = f_{2323} = f_{3232}$; |
| 23, m3 | 21 | 3 | $f_{1111} = f_{2222} = f_{3333}$; $f_{1122} = f_{1221} = f_{2121} = f_{2233} = f_{2332} = f_{3232} = f_{3311} = f_{3113} = f_{1313}$; $f_{2211} = f_{2112} = f_{1212} = f_{3322} = f_{3223} = f_{2323} = f_{1133} = f_{1331} = f_{3131}$; |
| $\bar{6}m2$, 622, 6mm, 6/mmm | 21 | 4 | $f_{1111} = f_{2222}$; $f_{3333}$; $f_{1122} = f_{1221} = f_{2121} = f_{2211} = f_{2112} = f_{1212} = \frac{f_{1111}}{3}$; $f_{1133} = f_{1331} = f_{3131} = f_{2233} = f_{2332} = f_{3232}$; $f_{1313} = f_{3113} = f_{3311} = f_{2323} = f_{3223} = f_{3322}$; |
| 6, $\bar{6}$, 6/m | 35 | 6 | $f_{1111} = f_{2222}$; $f_{3333}$; $f_{1121} = -f_{2212}$, $f_{1222} = f_{2122} = f_{2221} = \frac{f_{1121}}{3} = -f_{1112} = -f_{1211} = -f_{2111}$; $f_{1122} = f_{1221} = f_{2121} = f_{2211} = f_{2112} = f_{1212} = \frac{f_{1111}}{3}$; $f_{1133} = f_{1331} = f_{3131} = f_{2233} = f_{2332} = f_{3232}$; $f_{1313} = f_{3311} = f_{3113} = f_{3322} = f_{3223} = f_{2323}$; $f_{2313} = f_{3213} = f_{3312}$, $f_{1323} = f_{3123} = f_{3321} = -f_{2313}$; |
| 3, $\bar{3}$ | 67 | 10 | $f_{1111}=f_{2222};\ f_{3333};$<br>$f_{1121}=-f_{2212},$<br>$f_{1112}=f_{1211}=f_{2111}=-\frac{f_{1121}}{3}=-f_{1222}=-f_{2122}=-f_{2221};$<br>$f_{1122}=f_{1221}=f_{2121}=f_{2211}=f_{2112}=f_{1212}=\frac{f_{1111}}{3}$<br>$f_{1113}=f_{1311}=f_{3111};$<br>$f_{1223}=f_{1322}=f_{2123}=f_{2321}=f_{3122}=f_{3212}$<br>$=f_{2213}=f_{2312}=f_{3221}=-f_{1113};$<br>$f_{1123}=f_{1321}=f_{3121}=f_{1213}=f_{1312}=f_{2113}=f_{2311}=f_{3112}=f_{3211};$<br>$f_{2223}=f_{2322}=f_{3222}=-f_{1123};$<br>$f_{1131}=-f_{1232}=-f_{2132}=-f_{2231};$<br>$f_{1132}=f_{1231}=f_{2131}=-f_{2232};$<br>$f_{1133}=f_{2233}=f_{1331}=f_{2332}=f_{3131}=f_{3232};$<br>$f_{1313}=f_{3113}=f_{3311}=f_{2323}=f_{3223}=f_{3322};$<br>$f_{1323}=f_{3123}=f_{3321}=-f_{2313}=-f_{3213}=-f_{3312};$ |
| 32, 3m, $\bar{3}m$ | 37 | 6 | $f_{1111}=f_{2222};\ f_{3333};$<br>$f_{1122}=f_{1221}=f_{2121}=f_{2211}=f_{2112}=f_{1212}=\frac{f_{1111}}{3};$<br>$f_{1113}=f_{1311}=f_{3111}=-f_{1223}=-f_{1322}=-f_{2123}=-f_{2321}=-f_{3122}$<br>$=-f_{3221}=-f_{2213}=-f_{2312}=-f_{3212};$<br>$f_{1131}, f_{1232}=f_{2132}=f_{2231}=-f_{1131};$<br>$f_{1133}=f_{2233}=f_{1331}=f_{2332}=f_{3131}=f_{3232};$<br>$f_{1313}=f_{3113}=f_{3311}=f_{2323}=f_{3223}=f_{3322};$ |
| $\bar{4}\,2m$, 422, 4mm, 4/mmm | 21 | 5 | $f_{1111}=f_{2222};\ f_{3333};$<br>$f_{1122}=f_{1221}=f_{2121}=f_{2211}=f_{2112}=f_{1212};$<br>$f_{1133}=f_{1331}=f_{3131}=f_{2233}=f_{3232}=f_{2332};$<br>$f_{1313}=f_{3311}=f_{3113}=f_{2323}=f_{3322}=f_{3223};$ |
| 4,$\bar{4}$, 4/m | 35 | 8 | $f_{1111}=f_{2222};\ f_{3333};$<br>$f_{1122}=f_{1221}=f_{2121}=f_{2211}=f_{2112}=f_{1212};$<br>$f_{1133}=f_{1331}=f_{3131}=f_{2233}=f_{2332}=f_{3232};$<br>$f_{3311}=f_{3113}=f_{1313}=f_{3322}=f_{3223}=f_{2323};$<br>$f_{1121}=-f_{2212};$<br>$f_{1112}=f_{1211}=f_{2111}=-f_{1222}=-f_{2122}=-f_{2221};$<br>$f_{2313}=f_{3213}=f_{3312}=-f_{1323}=-f_{3123}=-f_{3321};$ |
| 222, mm2, mmm | 21 | 9 | $f_{1111},\ f_{2222}, f_{3333};$<br>$f_{1122}=f_{1221}=f_{2121};\ f_{2233}=f_{2332}=f_{3232};\ f_{3311}=f_{3113}=f_{1313};$<br>$f_{2211}=f_{2112}=f_{1212};\ f_{3322}=f_{3223}=f_{2323};\ f_{1133}=f_{1331}=f_{3131};$ |
| 2, m, 2/m | 41 | 16 | $f_{1111},\ f_{2222}, f_{3333};$<br>$f_{1122}=f_{1221}=f_{2121}; f_{2211}=f_{2112}=f_{1212};$<br>$f_{1112}=f_{1211}=f_{2111}; f_{1222}=f_{2122}=f_{2221};$<br>$f_{1133}=f_{1331}=f_{3131}; f_{1313}=f_{3113}=f_{3311};$<br>$f_{1233}=f_{2133}=f_{2331}=f_{1332}=f_{3132}=f_{3231};$<br>$f_{1323}=f_{3123}=f_{3321};\ f_{2313}=f_{3213}=f_{3312};$<br>$f_{2233}=f_{2332}=f_{3232};\ f_{2323}=f_{3223}=f_{3322};$<br>$f_{1121};\ f_{2212};$ |
| 1, $\bar{1}$ | 81 | 30 | $f_{1111}, f_{1112}=f_{1211}=f_{2111}, f_{1113}=f_{1311}=f_{3111};$<br>$f_{1121}, f_{1122}=f_{1221}=f_{2121}, f_{1123}=f_{1321}=f_{3121};$<br>$f_{1131}, f_{1132}=f_{1231}=f_{2131}, f_{1133}=f_{1331}=f_{3131};$<br>$f_{1212}=f_{2112}=f_{2211}, f_{1213}=f_{1312}=f_{3112}=f_{3211}=f_{2311}=f_{2113};$<br>$f_{1222}=f_{2122}=f_{2221}, f_{1223}=f_{1322}=f_{3122}=f_{3221}=f_{2321}=f_{2123};$<br>$f_{1232}=f_{2132}=f_{2231}, f_{1233}=f_{1332}=f_{3132}=f_{3231}=f_{2331}=f_{2133};$ |
| | | | $f_{1313}=f_{3113}=f_{3311}, f_{1323}=f_{3123}=f_{3321}, f_{1333}=f_{3133}=f_{3331};$<br>$f_{2212}, f_{2213}=f_{2312}=f_{3212};$<br>$f_{2222}, f_{2223}=f_{2322}=f_{3222};$<br>$f_{2232}, f_{2233}=f_{2332}=f_{3232};$<br>$f_{2313}=f_{3213}=f_{3312}, f_{2323}=f_{3223}=f_{3322}, f_{2333}=f_{3233}=f_{3332};$<br>$f_{3313}, f_{3323}, f_{3333}$ . |

### 3. Comparison of EIPS FET $\bar{f}_{ijkl}$ and HIPS FET $f_{ijkl}$ for different point groups

**Table AIII.** The comparison of EIPS FET $\bar{f}_{\mathrm{ijkl}}$ and HIPS FET $f_{ijkl}$ for different point groups

| | EIPS FET | | HIPS FET | | |
|---|---|---|---|---|---|
| point group | non-zero | independent | non-zero | independent | "difference" between two tensors |
| m3m, 432<br>$\bar{4}$3m | 21 | 3 | 21 | 2 | $\bar{f}_{1122}\neq\bar{f}_{1212}$; $f_{1122}=f_{1212}$ |
| 23, m3 | 21 | 5 | 21 | 3 | $\bar{f}_{1122}\neq\bar{f}_{1221}, \bar{f}_{2211}\neq\bar{f}_{1212}$; $f_{1122}=f_{1221}, f_{2211}=f_{1212}$ |
| $\bar{6}m2$,<br>622,<br>6mm,<br>6/mmm | 21 | 7 | 21 | 4 | $\bar{f}_{1212}=\bar{f}_{2112}=\bar{f}_{1221}=\bar{f}_{2121}=\frac{\bar{f}_{1111}}{2}-\frac{\bar{f}_{1122}}{2}\neq\bar{f}_{1122}$<br>$\bar{f}_{1133}\neq\bar{f}_{1331}, \bar{f}_{2233}\neq\bar{f}_{2332},$<br>$f_{1221}=f_{2121}=f_{2211}=f_{2112}=f_{1212}=\frac{f_{1111}}{3}=f_{1122}$<br>$f_{1133}=f_{1331}, f_{2233}=f_{2332},$ |
| 6, $\bar{6}$, 6/m | 39 | 12 | 35 | 6 | $\bar{f}_{1122}\neq\bar{f}_{1221}\neq\frac{\bar{f}_{1111}}{3}; \bar{f}_{1133}\neq\bar{f}_{1331}; \bar{f}_{3311}\neq\bar{f}_{1313}; \bar{f}_{3312}\neq\bar{f}_{3213}$<br>$\bar{f}_{1332}=\bar{f}_{3132}=-\bar{f}_{2331}=-\bar{f}_{3231}\neq 0$<br>$f_{1122}=f_{1221}=\frac{f_{1111}}{3}; f_{1133}=f_{1331}; f_{1313}=f_{3311}; f_{3312}=f_{3213}$<br>$f_{1332}=f_{3132}=f_{2331}=f_{3231}=0$ |
| 3, $\bar{3}$ | 71 | 18 | 67 | 10 | $\bar{f}_{1112}\neq\bar{f}_{1211}=-\frac{\bar{f}_{1112}}{2}-\frac{\bar{f}_{1121}}{2}; \bar{f}_{1113}\neq\bar{f}_{1311};$<br>$\bar{f}_{1122}\neq\bar{f}_{1221}=\frac{\bar{f}_{1111}}{2}-\frac{\bar{f}_{1122}}{2}; \bar{f}_{1123}\neq\bar{f}_{1321};$<br>$\bar{f}_{1133}\neq\bar{f}_{1331}; \bar{f}_{3311}\neq\bar{f}_{3113}; \bar{f}_{3123}\neq\bar{f}_{3321}$<br>$\bar{f}_{1332}=\bar{f}_{3132}=-\bar{f}_{2331}=-\bar{f}_{3231}\neq 0;$<br>$f_{1112}=f_{1211}=-\frac{f_{1121}}{3}; f_{1113}=f_{1311};$<br>$f_{1122}=f_{1221}=\frac{f_{1111}}{3}; f_{1123}=f_{1321}$<br>$f_{1133}=f_{1331}; f_{3311}=f_{3113}; f_{3123}=f_{3321}$<br>$f_{1332}=f_{3132}=f_{2331}=f_{3231}=0$ |
| 32, 3m,<br>$\bar{3}m$ | 37 | 10 | 37 | 6 | $\bar{f}_{1122}\neq\bar{f}_{1221}=\frac{\bar{f}_{1111}}{2}-\frac{\bar{f}_{1122}}{2}; \bar{f}_{1133}\neq\bar{f}_{1331};$<br>$\bar{f}_{1113}\neq\bar{f}_{1311}; \bar{f}_{1313}\neq\bar{f}_{3311};$<br>$f_{1122}=f_{1221}=\frac{f_{1111}}{3}; f_{1133}=f_{1331};$<br>$f_{1113}=f_{1311}; f_{1313}=f_{3311}$ |
| $\bar{4}\,2m$, 422,<br>4mm,<br>4/mmm | 21 | 8 | 21 | 5 | $\bar{f}_{1122}\neq\bar{f}_{1221}, \bar{f}_{1133}\neq\bar{f}_{1331}, \bar{f}_{2233}\neq\bar{f}_{2332},$<br>$f_{1122}=f_{1221}, f_{1133}=f_{1331}, f_{2233}=f_{2332}.$ |
| 4, $\bar{4}$, 4/m | 39 | 14 | 35 | 8 | $\bar{f}_{1122} \neq \bar{f}_{1221}$; $\bar{f}_{1133} \neq \bar{f}_{1331}$; $\bar{f}_{3311} \neq \bar{f}_{1313}$;<br>$\bar{f}_{1211} \neq \bar{f}_{1112}$; $\bar{f}_{3213} \neq \bar{f}_{3312}$<br>$\bar{f}_{1332} = \bar{f}_{3132} = -\bar{f}_{2331} = -\bar{f}_{3231} \neq 0$;<br>$f_{1122} = f_{1221}$; $f_{1133} = f_{1331}$; $f_{3311} = f_{1313}$;<br>$f_{1112} = f_{1211}$; $f_{3213} = f_{3312}$<br>$f_{1332} = f_{3132} = f_{2331} = f_{3231} = 0$ |
| 222, mm2, mmm | 21 | 15 | 21 | 9 | $\bar{f}_{1122} \neq \bar{f}_{1221}$, $\bar{f}_{1133} \neq \bar{f}_{1331}$, $\bar{f}_{2233} \neq \bar{f}_{2332}$,<br>$\bar{f}_{2211} \neq \bar{f}_{2112}$, $\bar{f}_{3311} \neq \bar{f}_{3113}$, $\bar{f}_{3322} \neq \bar{f}_{3223}$;<br>$f_{1122} = f_{1221}$, $f_{1133} = f_{1331}$, $f_{2233} = f_{2332}$,<br>$f_{2211} = f_{2112}$, $f_{3311} = f_{3113}$, $f_{3322} = f_{3223}$. |
| 2, m, 2/m | 41 | 28 | 41 | 16 | $\bar{f}_{1112} \neq \bar{f}_{1211}$; $\bar{f}_{1122} \neq \bar{f}_{1221}$; $\bar{f}_{1133} \neq \bar{f}_{1331}$; $\bar{f}_{1222} \neq \bar{f}_{2221}$;<br>$\bar{f}_{1233} \neq \bar{f}_{1332} \neq \bar{f}_{2331} \neq \bar{f}_{1233}$; $\bar{f}_{1313} \neq \bar{f}_{3311}$; $\bar{f}_{1323} \neq \bar{f}_{3321}$;<br>$\bar{f}_{2211} \neq \bar{f}_{2112}$; $\bar{f}_{2233} \neq \bar{f}_{2332}$; $\bar{f}_{2313} \neq \bar{f}_{3312}$; $\bar{f}_{2323} \neq \bar{f}_{3322}$;<br><br>$f_{1112} = f_{1211}$; $f_{1122} = f_{1221}$; $f_{1133} = f_{1331}$; $f_{1222} = f_{2122}$;<br>$f_{1233} = f_{2331} = f_{1332}$; $f_{1313} = f_{3311}$; $f_{1323} = f_{3321}$;<br>$f_{2211} = f_{2112}$; $f_{2233} = f_{2332}$; $f_{2313} = f_{3312}$; ; $f_{2323} = f_{3322}$; |
| 1, $\bar{1}$ | 81 | 54 | 81 | 30 | $\bar{f}_{1112} \neq \bar{f}_{1211}$, $\bar{f}_{1113} \neq \bar{f}_{1311}$, $\bar{f}_{1122} \neq \bar{f}_{1221}$, $\bar{f}_{1123} \neq \bar{f}_{1321}$,<br>$\bar{f}_{1132} \neq \bar{f}_{1231}$, $\bar{f}_{1133} \neq \bar{f}_{1331}$, $\bar{f}_{1212} \neq \bar{f}_{2211}$,<br>$\bar{f}_{1213} \neq \bar{f}_{1312}$, $\bar{f}_{1213} \neq \bar{f}_{2311}$, $\bar{f}_{1222} \neq \bar{f}_{2221}$, $\bar{f}_{1223} \neq \bar{f}_{1322}$,<br>$\bar{f}_{1223} \neq \bar{f}_{2321}$, $\bar{f}_{1232} \neq \bar{f}_{2231}$ $\bar{f}_{1233} \neq \bar{f}_{1332}$, $\bar{f}_{1233} \neq \bar{f}_{2331}$,<br>$\bar{f}_{1312} \neq \bar{f}_{2311}$, $\bar{f}_{1313} \neq \bar{f}_{3311}$, $\bar{f}_{1323} \neq \bar{f}_{3321}$, $\bar{f}_{1322} \neq \bar{f}_{2321}$<br>$\bar{f}_{1332} \neq \bar{f}_{2331}$, $\bar{f}_{1333} \neq \bar{f}_{3331}$,<br>$\bar{f}_{2213} \neq \bar{f}_{2312}$, $\bar{f}_{2223} \neq \bar{f}_{2322}$, $\bar{f}_{2233} \neq \bar{f}_{2332}$;<br>$\bar{f}_{2313} \neq \bar{f}_{3312}$, $\bar{f}_{2323} \neq \bar{f}_{3322}$, $\bar{f}_{2333} \neq \bar{f}_{3332}$<br><br>$f_{1112} = f_{1211}$, $f_{1113} = f_{1311}$, $f_{1122} = f_{1221}$, $f_{1123} = f_{1321}$;<br>$f_{1132} = f_{1231}$, $f_{1133} = f_{1331}$; $f_{1212} = f_{2211}$, $f_{1222} = f_{2221}$,<br>$f_{1213} = f_{1312} = f_{3112} = f_{3211} = f_{2311} = f_{2113}$, $f_{1232} = f_{2231}$,<br>$f_{1223} = f_{1322} = f_{3122} = f_{3221} = f_{2321} = f_{2123}$, $f_{1323} = f_{3321}$,<br>$f_{1233} = f_{1332} = f_{3132} = f_{3231} = f_{2331} = f_{2133}$, $f_{1333} = f_{3331}$,<br>$f_{1313} = f_{3311}$, $f_{2213} = f_{2312}$, $f_{2223} = f_{2322}$, $f_{2233} = f_{2332}$,<br>$f_{2313} = f_{3312}$, $f_{2323} = f_{3322}$, $f_{2333} = f_{3332}$ |

## Appendix A2

### Application of the flexosymmetry to the Lifshitz invariant

The application of the results from **Table AI** (**Table I** from the text) for **cubic (m3m, 432, 4'3m)** point symmetry group to the Lifshitz invariant (1) yields

$$\begin{aligned}\Delta F_{FL}^{c} = &-\left(f_{1111}u_{11} + f_{1122}\left(u_{22}+u_{33}\right)\right)\frac{\partial P_1}{\partial x_1} - \left(f_{1111}u_{22} + f_{1122}\left(u_{11}+u_{33}\right)\right)\frac{\partial P_2}{\partial x_2} \\ &-\left(f_{1111}u_{33} + f_{1122}\left(u_{22}+u_{11}\right)\right)\frac{\partial P_3}{\partial x_3} \\ &-2f_{1212}\left(u_{12}\left(\frac{\partial P_1}{\partial x_2}+\frac{\partial P_2}{\partial x_1}\right)+u_{13}\left(\frac{\partial P_1}{\partial x_3}+\frac{\partial P_3}{\partial x_1}\right)+u_{23}\left(\frac{\partial P_2}{\partial x_3}+\frac{\partial P_3}{\partial x_2}\right)\right)\end{aligned} \tag{A.4}$$

The application of the results from **Table AII** (**Table II** from the text) for **cubic (m3m, 432, 4'3m)** point symmetry group to the Lifshitz invariant (1) yields

$$\begin{aligned}\Delta F_{FL}^{c} = &-\left(f_{1111}u_{11} + f_{1122}\left(u_{22}+u_{33}\right)\right)\frac{\partial P_1}{\partial x_1} - \left(f_{1111}u_{22} + f_{1122}\left(u_{11}+u_{33}\right)\right)\frac{\partial P_2}{\partial x_2} \\ &-\left(f_{1111}u_{33} + f_{1122}\left(u_{22}+u_{11}\right)\right)\frac{\partial P_3}{\partial x_3} \\ &-2f_{1122}\left(u_{12}\left(\frac{\partial P_1}{\partial x_2}+\frac{\partial P_2}{\partial x_1}\right)+u_{13}\left(\frac{\partial P_1}{\partial x_3}+\frac{\partial P_3}{\partial x_1}\right)+u_{23}\left(\frac{\partial P_2}{\partial x_3}+\frac{\partial P_3}{\partial x_2}\right)\right)\end{aligned} \tag{A.5}$$

The application of the results from **Table AI** (**Table I** from the text) for **tetragonal (4'2m, 422, 4mm, 4/mmm)** point symmetry group to the Lifshitz invariant (1) yields

$$\begin{aligned}\Delta F_{FL}^{t} = &-f_{1111}\left(u_{11}\frac{\partial P_1}{\partial x_1}+u_{22}\frac{\partial P_2}{\partial x_2}\right)-f_{3333}u_{33}\frac{\partial P_3}{\partial x_3}- \\ &f_{1122}\left(u_{11}\frac{\partial P_2}{\partial x_2}+u_{22}\frac{\partial P_1}{\partial x_1}\right)-f_{1133}\left(u_{11}\frac{\partial P_3}{\partial x_3}+u_{22}\frac{\partial P_3}{\partial x_3}\right)-f_{3311}\left(u_{33}\frac{\partial P_1}{\partial x_1}+u_{33}\frac{\partial P_2}{\partial x_2}\right) \\ &-2f_{1212}u_{12}\left(\frac{\partial P_2}{\partial x_1}+\frac{\partial P_1}{\partial x_2}\right)-2f_{1331}\left(u_{13}\frac{\partial P_3}{\partial x_1}+u_{23}\frac{\partial P_3}{\partial x_2}\right)-2f_{1313}\left(u_{13}\frac{\partial P_1}{\partial x_3}+u_{23}\frac{\partial P_2}{\partial x_3}\right)\end{aligned} \tag{A.6}$$

The application of the results from **Table AII** (**Table II** from the text) for **tetragonal (4'2m, 422, 4mm, 4/mmm)** point symmetry group to the Lifshitz invariant (1) yields

$$\Delta F_{FL}^{t} = -f_{1111}\left(u_{11}\frac{\partial P_1}{\partial x_1} + u_{22}\frac{\partial P_2}{\partial x_2}\right) - f_{3333}u_{33}\frac{\partial P_3}{\partial x_3}$$
$$-f_{1122}\left(u_{11}\frac{\partial P_2}{\partial x_2} + u_{22}\frac{\partial P_1}{\partial x_1} + 2u_{12}\left(\frac{\partial P_2}{\partial x_1} + \frac{\partial P_1}{\partial x_2}\right)\right)$$
$$-f_{1133}\left(u_{11}\frac{\partial P_3}{\partial x_3} + u_{22}\frac{\partial P_3}{\partial x_3} + 2u_{13}\frac{\partial P_3}{\partial x_1} + 2u_{23}\frac{\partial P_3}{\partial x_2}\right)$$
$$-f_{3311}\left(u_{33}\frac{\partial P_1}{\partial x_1} + u_{33}\frac{\partial P_2}{\partial x_2} + 2u_{13}\frac{\partial P_1}{\partial x_3} + 2u_{23}\frac{\partial P_2}{\partial x_3}\right) \quad (A.7)$$

The application of the results from **Table AII** (**Table II** from the text) for **orthorhombic (222, mm2, mmm)** point symmetry group to the Lifshitz invariant (1) yields

$$\Delta F_{FL}^{o} = -f_{1111}u_{11}\frac{\partial P_1}{\partial x_1} - f_{2222}u_{22}\frac{\partial P_2}{\partial x_2} - f_{3333}u_{33}\frac{\partial P_3}{\partial x_3}$$
$$-f_{1122}\left(u_{11}\frac{\partial P_2}{\partial x_2} + 2u_{12}\frac{\partial P_2}{\partial x_1}\right) - f_{2211}\left(u_{22}\frac{\partial P_1}{\partial x_1} + 2u_{12}\frac{\partial P_1}{\partial x_2}\right)$$
$$-f_{1133}\left(u_{11}\frac{\partial P_3}{\partial x_3} + 2u_{13}\frac{\partial P_3}{\partial x_1}\right) - f_{2233}\left(u_{22}\frac{\partial P_3}{\partial x_3} + 2u_{23}\frac{\partial P_3}{\partial x_2}\right)$$
$$-f_{3322}\left(u_{33}\frac{\partial P_2}{\partial x_2} + 2u_{23}\frac{\partial P_2}{\partial x_3}\right) - f_{3311}\left(u_{33}\frac{\partial P_1}{\partial x_1} + 2u_{13}\frac{\partial P_1}{\partial x_3}\right) \quad (A.8)$$

The expressions for mmm symmetry without taking into account the full index-permutation symmetry (2) is too cumbersome and we omit it for the sake of brevity.

One could visualize the invariants (A.4)-(A.8) allowing for the point group symmetry and full index-permutation symmetry (2), if all values of the flexoelectric and elastic tensors are well-known either from experiment or from *ab initio* calculations.

# Appendix B. Effects of Point Symmetry

Consider an infinite crystal under an inhomogeneous stress or/and inhomogeneous electric field. The increment in free energy due to the flexoelectric effect, including the strain-induced polarization and polarization-induced strain, is given by Lifshitz invariant.[4] Under assumptions of the continuous model, the corresponding lattice sum is approximated by the integral (1). Integration by parts transforms it to a more symmetric form[7] which, per unit cell, reads as

$$\Delta F = -\frac{1}{2}\sum_{i,j,k,l}\left(f_{ijkl}+f_{jikl}\right)\frac{\partial U_i}{\partial x_j}\frac{\partial P_k}{\partial x_l}. \qquad ( .1)$$

Here **U** – is the displacement vector of the unit cell deformation. Other symbols are the same as in (1). The second-rank tensor $\partial U_i/\partial x_j$ can be presented as the sum of its symmetric and antisymmetric parts, $\partial U_i/\partial x_j = u_{ij} + r_{ij}$, with

$$u_{ij} = \frac{1}{2}\left(\frac{\partial U_i}{\partial x_j}+\frac{\partial U_j}{\partial x_i}\right), \qquad r_{ij} = \frac{1}{2}\left(\frac{\partial U_i}{\partial x_j}-\frac{\partial U_j}{\partial x_i}\right) \qquad ( .2)$$

For the tensor $f_{ijkl}$ in (B.1), in general, the number of its components is $3^4 = 81$. As mentioned in Section II, not all of them are independent. The number of independent coefficients depends on the point symmetry of the crystal, as well as on the permutational symmetry of the indices in $f_{ijkl}$, its hidden symmetry[7]. In what follows, the role of the point symmetry is discussed. For simplicity, the antisymmetric part of $\partial U_i/\partial x_j$ is set to zero, $r_{ij} = 0$. This gives $\partial U_i/\partial x_j = u_{ij}$ meaning the tensor $\partial U_i/\partial x_j$ is symmetric.

For simplicity, let the crystal lattice be of a symmorphic space symmetry, meaning there are no screw axes and no gliding planes of symmetry. Low-symmetry effects of crystal surface, domain walls, and other defects are neglected. The free energy $\Delta F$ of (B.1) is a scalar of the crystal symmetry group. Therefore, the right-hand side of (B.1) includes just scalar convolutions of the two second-rank tensors, $\partial U_i/\partial x_j$ and $\partial P_i/\partial x_j$.

By their transformation properties, components of both tensors form the basis of reducible representations, which can be decomposed into irreducible ones. The corresponding symmetry-adapted combinations are

$$\frac{\partial U_i}{\partial x_j} = \sum_{\Gamma\gamma}\langle\Gamma\gamma|\Gamma_1 i\Gamma_2 j\rangle\left(\frac{\partial U}{\partial x}\right)_{\Gamma\gamma}, \qquad \frac{\partial P_k}{\partial x_l} = \sum_{\Gamma\gamma}\langle\Gamma\gamma|\Gamma_1 k\Gamma_2 l\rangle\left(\frac{\partial P}{\partial x}\right)_{\Gamma\gamma} \qquad ( .3)$$

Here, $\langle\Gamma\gamma|\Gamma_1 i\Gamma_2 j\rangle$ are Clebsch – Gordan coefficients, Γ and γ denote irreducible representations and their rows present in the product $\Gamma_1 \times \Gamma_2$ of the two representations $\Gamma_1$ and $\Gamma_2$. As both $U_i$ and $\partial/\partial x_j$ transform as vectors, $\Gamma_2$ and $\Gamma_1$, are vector representations of the corresponding point group. For the symmetric tensor $\partial U_i/\partial x_j = u_{ij}$, we only include those Γ that are present in the symmetric square $[\Gamma_1^2]$. As the matrix of Clebsch – Gordan coefficients is orthogonal, the inverse transformation follows from (B.2) by transposition:

$$\left(\frac{\partial U}{\partial x}\right)_{\Gamma\gamma} = \sum_{i,j}\langle\Gamma_1 i\Gamma_2 j|\Gamma\gamma\rangle\frac{\partial U_i}{\partial x_j}, \qquad \left(\frac{\partial P}{\partial x}\right)_{\Gamma\gamma} = \sum_{i,j}\langle\Gamma_1 i\Gamma_2 j|\Gamma\gamma\rangle\frac{\partial P_i}{\partial x_j} \qquad ( .4)$$

Substituting (B.3) into (B.1), we transform it into the a bilinear form of the first-rank irreducible tensors, $(\partial P/\partial x)_{\Gamma\gamma}$ and $(\partial U/\partial x)_{\Gamma\gamma}$, so the products $(\partial U/\partial x)_{\Gamma\gamma}(\partial P/\partial x)_{\bar{\Gamma}\bar{\gamma}}$ are components of the second-rank tensor. As mentioned above, the resultant sum is a scalar of the crystal-lattice point group. Therefore, in this sum we keep just scalar convolutions belonging to the totally symmetric representation of the point group. The latter is present in symmetric squares $[\Gamma^2]$ only, once in each. Therefore, there are as many scalar convolutions as different squares $[\Gamma^2]$ can be formed. In what follows below, a few examples are presented for several point groups.

(*a*) *Cubic symmetry group m3m.* Vectors belong to the irreducible representation $T_{1u}$. Therefore, in (B.2) we have $\Gamma_1 = T_{1u}$ and $\Gamma_2 = T_{1u}$. Correspondingly, in (B.3) and (B.4), Γs are the representations present in the product $T_{1u} \times T_{1u} = A_{1g} + E_g + T_{1g} + T_{2g}$. As mentioned above, for $\partial U_i/\partial x_j$ only the symmetric square is included, $[T_{1u}{}^2] = A_{1g} + E_g + T_{2g}$. Plugging the corresponding Clebsch–Gordan coefficients into (B.4) and substituting $\partial U_i/\partial x_j = u_{ij}$, we come to symmetry-adapted components,[1]

$$A_{1g}: \quad u_A = \frac{1}{\sqrt{3}}\left(u_{11}+u_{22}+u_{33}\right),$$

$$E_g: \begin{cases} u_\theta = \dfrac{1}{\sqrt{6}}\left(u_{11}+u_{22}-2u_{33}\right) \\ u_\varepsilon = \dfrac{1}{\sqrt{2}}\left(-u_{11}+u_{22}\right) \end{cases}, \qquad T_{2g}: \begin{cases} u_\xi = \dfrac{1}{\sqrt{2}}\left(u_{23}+u_{32}\right)=\sqrt{2}u_{23} \\ u_\eta = \dfrac{1}{\sqrt{2}}\left(u_{13}+u_{31}\right)=\sqrt{2}u_{13} \\ u_\zeta = \dfrac{1}{\sqrt{2}}\left(u_{12}+u_{21}\right)=\sqrt{2}u_{12} \end{cases} \qquad ( .5)$$

Similarly, for $(\partial P/\partial x)_{\Gamma\gamma}$,

$$A_{1g}: \left(\frac{\partial P}{\partial x}\right)_A = \frac{1}{\sqrt{3}}\left(\frac{\partial P_1}{\partial x_1}+\frac{\partial P_2}{\partial x_2}+\frac{\partial P_3}{\partial x_3}\right), \quad E_g: \begin{cases} \left(\dfrac{\partial P}{\partial x}\right)_\theta = \dfrac{1}{\sqrt{6}}\left(\dfrac{\partial P_1}{\partial x_1}+\dfrac{\partial P_2}{\partial x_2}-2\dfrac{\partial P_3}{\partial x_3}\right) \\ \left(\dfrac{\partial P}{\partial x}\right)_\varepsilon = \dfrac{1}{\sqrt{2}}\left(-\dfrac{\partial P_1}{\partial x_1}+\dfrac{\partial P_2}{\partial x_2}\right) \end{cases},$$

$$T_{1g}: \begin{cases} \left(\dfrac{\partial P}{\partial x}\right)_\xi = \dfrac{1}{\sqrt{2}}\left(\dfrac{\partial P_2}{\partial x_3}-\dfrac{\partial P_3}{\partial x_2}\right) \\ \left(\dfrac{\partial P}{\partial x}\right)_\eta = \dfrac{1}{\sqrt{2}}\left(\dfrac{\partial P_1}{\partial x_3}-\dfrac{\partial P_3}{\partial x_1}\right), \\ \left(\dfrac{\partial P}{\partial x}\right)_\zeta = \dfrac{1}{\sqrt{2}}\left(\dfrac{\partial P_1}{\partial x_2}-\dfrac{\partial P_2}{\partial x_1}\right) \end{cases} \qquad T_{2g}: \begin{cases} \left(\dfrac{\partial P}{\partial x}\right)_\xi = \dfrac{1}{\sqrt{2}}\left(\dfrac{\partial P_2}{\partial x_3}+\dfrac{\partial P_3}{\partial x_2}\right) \\ \left(\dfrac{\partial P}{\partial x}\right)_\eta = \dfrac{1}{\sqrt{2}}\left(\dfrac{\partial P_1}{\partial x_3}+\dfrac{\partial P_3}{\partial x_1}\right) \\ \left(\dfrac{\partial P}{\partial x}\right)_\zeta = \dfrac{1}{\sqrt{2}}\left(\dfrac{\partial P_1}{\partial x_2}+\dfrac{\partial P_2}{\partial x_1}\right) \end{cases} \qquad ( .6)$$

---

[1]For irreducible representations of cubic group, we use the real basis set widely used in literature. The corresponding transformation properties are: $E_g\{\theta \sim x^2 + y^2 - 2z^2, \varepsilon \sim -x^2 + y^2\}$, $T_{1u}\{x, y, z\}$, and $T_{2g}\{\xi \sim yz, \eta \sim xz, z \sim xy\}$.

Therefore, in (B.1) scalars come from the products $(A_{1g} + E_g + T_{2g}) \times (A_{1g} + E_g + T_{1g} + T_{2g})$, where the first factor stands for strain and the second one corresponds to $\partial P_k/\partial x_l$. In this product the totally symmetric representations $A_{1g}$ are present in symmetric squares only, $[A_{1g}^2]$, $[E_g^2]$ and $[T_{2g}^2]$. Correspondingly, (B.1) includes three scalars,

$$\begin{aligned} \Delta F = & -f\left[A_{1g}^2\right] u_A \left(\frac{\partial P}{\partial x}\right)_A - \frac{1}{\sqrt{2}} f\left[E_g^2\right]\left[u_\theta \left(\frac{\partial P}{\partial x}\right)_\theta + u_\varepsilon \left(\frac{\partial P}{\partial x}\right)_\varepsilon\right] \\ & - \frac{1}{\sqrt{3}} f\left[T_{2g}^2\right]\left[u_\xi \left(\frac{\partial P}{\partial x}\right)_\xi + u_\eta \left(\frac{\partial P}{\partial x}\right)_\eta + u_\zeta \left(\frac{\partial P}{\partial x}\right)_\zeta\right] \end{aligned} \qquad ( .7)$$

with the following three independent combinations of the flexoelectric coupling,

$$\begin{aligned} f\left[A_{1g}^2\right] &= f_{AA} = \frac{1}{3}\left(f_{1111} + f_{2222} + f_{3333} + f_{1122} + f_{1133} + f_{2211} + f_{2233} + f_{3311} + f_{3322}\right), \\ f\left[E_g^2\right] &= \frac{1}{\sqrt{2}}\left(f_{\theta\theta} + f_{\varepsilon\varepsilon}\right) \\ &= \frac{1}{3\sqrt{2}}\left(2f_{1111} + 2f_{2222} + 2f_{3333} - f_{1122} - f_{1133} - f_{2211} - f_{2233} - f_{3311} - f_{3322}\right), \\ f\left[T_{2g}^2\right] &= \frac{1}{\sqrt{3}}\left(f_{\xi\xi} + f_{\eta\eta} + f_{\zeta\zeta}\right) \\ &= \frac{1}{2\sqrt{3}}\left(f_{2323} + f_{2332} + f_{3223} + f_{3232} + f_{1313} + f_{1331} + f_{3113} + f_{3131} + f_{1212} + f_{2112} + f_{1221} + f_{2121}\right) \end{aligned} \qquad ( .8)$$

Thus, in a cubic crystal, of the 21 nonzero components of $f_{ijkl}$ in (B.1), only three combinations, $f[A_{1g}^2]$, $f[E_g^2]$ and $f[T_{2g}^2]$ of (B.8), are independent parameters. In addition to the outlined effect of point symmetry, the flexoelectric tensor, as it follows from its definition, has hidden symmetry with respect to the indeces,[7] $f_{ijkl} = f_{jikl}$ and $f_{ijkl} = f_{ilkj}$. Applying the relations listed in Table II, we find

$$f\left[A_{1g}^2\right] = f_{1111} + 2f_{1122}, \quad f\left[E_g^2\right] = \sqrt{2}\left(f_{1111} - f_{1122}\right), \quad f\left[T_{2g}^2\right] = 2\sqrt{3} f_{1122} \qquad ( .9)$$

As one can see in (B.9), the three constants of flexoelectric coupling, $f[A_{1g}^2]$, $f[E_g^2]$ and $f[T_{2g}^2]$, are expressed in terms of just two components, $f_{1111}$ and $f_{1122}$. This means, with the hidden symmetry included, the three coupling constants are not independent and can be expressed in terms of one another. Indeed, from (B.9) we find

$$2f\left[A_{1g}^2\right] = \sqrt{2} f\left[E_g^2\right] + \sqrt{3} f\left[T_{2g}^2\right]$$

Therefore, the hidden symmetry reduces the number of independent coupling constants from three to just two. Substituting (B.5), (B.6) and (B.9) into (B.7) we come to Eq.(A.5) of the Appendix A2.

(*b*) *Tetragonal symmetry group 4/mmm.* Vectors belong to irreducible representations $A_{2u}$ and $E_u$. Therefore, in (B.3) we have $\Gamma_1 = A_{2u}, E_u$ and $\Gamma_2 = A_{2u}, E_u$. Correspondingly, in (B.4) irreducible representations $\Gamma$ are the ones present in $(A_{2u} + E_u) \times (A_{2u} + E_u) = A_{2u} \times A_{2u} + 2A_{2u} \times E_u + E_u \times E_u = 2A_{1g} + A_{2g} + B_{1g} + B_{2g} + 2E_g$. Selecting scalar convolutions included in the products $(2A_{1g} + A_{2g}$

+ $B_{1g}$ + $B_{2g}$ + $2E_g$) × ($2A_{1g}$ + $A_{2g}$ + $B_{1g}$ + $B_{2g}$ + $2E_g$), where the first factor describes the transformation properties of the strain tensor, and the second relates to the tensor $\partial P_k/\partial x_l$. As above, due to the symmetric property of strain, in the first factor the antisymmetric representation $A_{2g}$ should be skipped. Hence, we come to scalars included in the product ($2A_{1g}$ + $B_{1g}$ + $B_{2g}$ + $2E_g$) × ($2A_{1g}$ + $A_{2g}$ + $B_{1g}$ + $B_{2g}$ + $2E_g$). Totally-symmetric representation $A_{1g}$ occurs in symmetric squares $[\Gamma^2]$ only. Therefore, we keep just the "diagonal" products, $4[A_{1g}{}^2]$ + $[B_{1g}{}^2]$ + $[B_{2g}{}^2]$ + $4[E_g{}^2]$, a total of 10 terms. This implies existence of ten scalars of the following types: $[B_{1g}{}^2]$, $[B_{2g}{}^2]$, and $[A_{1g}{}^2]_{mn}$, $[E_g{}^2]_{mn}$, with $m$, $n$ = 1,2. Hence, in the tetragonal group, the Lifshitz invariant includes 10 independent parameters $f([\Gamma^2])$:

$$\begin{aligned}\Delta F = &\sum_{m,n=1,2} f\left(A_{1g}\times A_{1g}\right)_{mn}\left(\frac{\partial U}{\partial x}\right)^{(m)}_{A_{1g}}\left(\frac{\partial P}{\partial x}\right)^{(n)}_{A_{1g}} + f\left[B_{1g}^2\right]\left(\frac{\partial U}{\partial x}\right)_{B_{1g}}\left(\frac{\partial P}{\partial x}\right)_{B_{1g}}\\ &+ f\left[B_{2g}^2\right]\left(\frac{\partial U}{\partial x}\right)_{B_{2g}}\left(\frac{\partial P}{\partial x}\right)_{B_{2g}} + \sum_{m,n=1,2} f\left(E_g\times E_g\right)_{mn}\left(\frac{\partial U}{\partial x}\right)^{(m)}_{E_g}\left(\frac{\partial P}{\partial x}\right)^{(n)}_{E_g}\end{aligned} \tag{.10}$$

Plugging Clebsch – Gordan coefficients in (B.4), we find the irreducible tensors of strain and polarization, and, correspondingly, all the scalars in (B.10). Components of the irreducible tensor of strain are:

$$\begin{aligned}&\left(\frac{\partial U}{\partial x}\right)^{(1)}_{A_{1g}} = u_{33},\quad \left(\frac{\partial U}{\partial x}\right)^{(2)}_{A_{1g}} = \frac{1}{\sqrt{2}}\left(u_{11}+u_{22}\right),\quad \left(\frac{\partial U}{\partial x}\right)_{B_{1g}} = \frac{1}{\sqrt{2}}\left(u_{11}-u_{22}\right),\quad \left(\frac{\partial U}{\partial x}\right)_{B_{2g}} = u_{12},\\ &\left(\frac{\partial U}{\partial x}\right)^{(1)}_{E_g x} = u_{13},\quad \left(\frac{\partial U}{\partial x}\right)^{(1)}_{E_g y} = u_{23},\quad \left(\frac{\partial U}{\partial x}\right)^{(2)}_{E_g x} = u_{13},\quad \left(\frac{\partial U}{\partial x}\right)^{(2)}_{E_g y} = u_{23}\end{aligned} \tag{.11}$$

Note the coincidence of the components $\left(\frac{\partial U}{\partial x}\right)^{(1)}_{E_g g}$ and $\left(\frac{\partial U}{\partial x}\right)^{(2)}_{E_g g}$. This is due to the above imposed restriction $r_{ij} = 0$ [see the comment immediately after (B.2)]. Similarly, for $\partial P_k/\partial x_l$ its irreducible components are:

$$\begin{aligned}&\left(\frac{\partial P}{\partial x}\right)^{(1)}_{A_{1g}} = \frac{\partial P_3}{\partial x_3},\quad \left(\frac{\partial P}{\partial x}\right)^{(2)}_{A_{1g}} = \frac{1}{\sqrt{2}}\left(\frac{\partial P_1}{\partial x_1}+\frac{\partial P_2}{\partial x_2}\right),\quad \left(\frac{\partial P}{\partial x}\right)_{B_{1g}} = \frac{1}{\sqrt{2}}\left(\frac{\partial P_1}{\partial x_1}-\frac{\partial P_2}{\partial x_2}\right),\\ &\left(\frac{\partial U}{\partial x}\right)_{B_{2g}} = \frac{1}{\sqrt{2}}\left(\frac{\partial P_1}{\partial x_2}+\frac{\partial P_2}{\partial x_1}\right),\quad \left(\frac{\partial P}{\partial x}\right)^{(1)}_{E_g x} = \frac{\partial P_1}{\partial x_3},\quad \left(\frac{\partial P}{\partial x}\right)^{(1)}_{E_g y} = \frac{\partial P_2}{\partial x_3},\quad \left(\frac{\partial P}{\partial x}\right)^{(2)}_{E_g x} = \frac{\partial P_3}{\partial x_1},\quad \left(\frac{\partial P}{\partial x}\right)^{(2)}_{E_g y} = \frac{\partial P_3}{\partial x_2}\end{aligned} \tag{.12}$$

Finally, here are the corresponding independent combinations of the tensor of flexoelectric coupling. As mentioned above, there are ten of them:

$$
\begin{gathered}
f\left(\left[A_{1g}^{2}\right]_{11}\right)=f_{3333},\quad f\left(\left[A_{1g}^{2}\right]_{12}\right)=\frac{1}{\sqrt{2}}\left(f_{3311}+f_{3322}\right),\quad f\left(\left[A_{1g}^{2}\right]_{21}\right)=\frac{1}{\sqrt{2}}\left(f_{1133}+f_{2233}\right),\\
f\left(\left[A_{1g}^{2}\right]_{22}\right)=\frac{1}{2}\left(f_{1111}+f_{1122}+f_{2211}+f_{2222}\right),\quad f\left(\left[B_{1g}^{2}\right]\right)=\frac{1}{2}\left(f_{1111}-f_{1122}-f_{2211}+f_{2222}\right),\\
f\left(\left[B_{2g}^{2}\right]\right)=\frac{1}{\sqrt{2}}\left(f_{1212}+f_{1221}\right),\quad f\left(\left[E_{g}^{2}\right]_{11}\right)=\frac{1}{\sqrt{2}}\left(f_{1331}+f_{2332}\right),\quad f\left(\left[E_{g}^{2}\right]_{12}\right)=\frac{1}{\sqrt{2}}\left(f_{1313}+f_{2323}\right),\\
f\left(\left[E_{g}^{2}\right]_{21}\right)=\frac{1}{\sqrt{2}}\left(f_{1331}+f_{2332}\right),\quad f\left(\left[E_{g}^{2}\right]_{22}\right)=\frac{1}{\sqrt{2}}\left(f_{1313}+f_{2323}\right)
\end{gathered}
\qquad (\ .13)
$$

As above, due to the restriction $r_{ij}$ = 0, there are two coinciding coupling constants, $f([E_g^2]_{11}) = f([E_g^2]_{21})$ and $f([E_g^2]_{12}) = f([E_g^2]_{22})$. With these two constraints, there are just eight independent parameters. Hidden symmetry (see Table II in the main text) simplifies (B.13) as follows:

$$
\begin{gathered}
f\left(\left[A_{1g}^{2}\right]_{11}\right)=f_{3333},\; f\left(\left[A_{1g}^{2}\right]_{12}\right)=\sqrt{2}f_{3311},\; f\left(\left[A_{1g}^{2}\right]_{21}\right)=\sqrt{2}f_{1133},\; f\left(\left[A_{1g}^{2}\right]_{22}\right)=f_{1111}+f_{1122},\\
f\left(\left[B_{1g}^{2}\right]\right)=f_{1111}-f_{1122},\quad f\left(\left[B_{2g}^{2}\right]\right)=\sqrt{2}f_{1122},\quad f\left(\left[E_{g}^{2}\right]_{11}\right)=f\left(\left[E_{g}^{2}\right]_{21}\right)=\sqrt{2}f_{1133},\\
f\left(\left[E_{g}^{2}\right]_{12}\right)=f\left(\left[E_{g}^{2}\right]_{22}\right)=\sqrt{2}f_{3311}
\end{gathered}
\qquad (\ .14)
$$

Here, like in the cubic case, the eight constants of flexoelectric coupling are expressed in terms of just five components, $f_{1111}$, $f_{3333}$, $f_{3311}$, $f_{1133}$, and $f_{1122}$. This means, with the hidden symmetry included, the eight coupling constants are not independent and can be expressed in terms of one another,

$$
f\left(\left[E_{g}^{2}\right]_{12}\right)=f\left(\left[E_{g}^{2}\right]_{22}\right)=f\left(\left[A_{1g}^{2}\right]_{12}\right),\qquad f\left(\left[E_{g}^{2}\right]_{11}\right)=f\left(\left[E_{g}^{2}\right]_{21}\right)=f\left(\left[A_{1g}^{2}\right]_{21}\right)
$$

Therefore, the hidden symmetry reduces the number of independent coupling constants from eight to just five. Substituting (B.12), (B.13) and (B.14) into (B.10), we come to Eq.(A.7) of the Appendix A2.

(c) *Orthorhombic symmetry, the group mmm.* In this symmetry group, vectors belong to irreducible representations $B_{1u}$, $B_{2u}$ and $B_{3u}$. Therefore, for $(\partial P/\partial x)_\Gamma$ in (B.3) and (B.4), the irreducible representations $\Gamma_1$ and $\Gamma_2$ are the ones present in $(B_{1u} + B_{2u} + B_{3u}) \times (B_{1u} + B_{2u} + B_{3u}) = 3A_{1g} + 2B_{1g} + 2B_{2g} + 2B_{3g}$, the total of nine items. Plugging Clebsch – Gordan coefficients in (B.4), we find the nine first-rank irreducible tensors:

$$
\begin{gathered}
\left(\frac{\partial P}{\partial x}\right)_{A_{1g}}^{(1)}=\frac{\partial P_1}{\partial x_1},\quad \left(\frac{\partial P}{\partial x}\right)_{A_{1g}}^{(2)}=\frac{\partial P_2}{\partial x_2},\quad \left(\frac{\partial P}{\partial x}\right)_{A_{1g}}^{(3)}=\frac{\partial P_3}{\partial x_3},\\
\left(\frac{\partial P}{\partial x}\right)_{B_{1g}}^{(1)}=\frac{1}{\sqrt{2}}\left(\frac{\partial P_1}{\partial x_2}+\frac{\partial P_2}{\partial x_1}\right),\quad \left(\frac{\partial P}{\partial x}\right)_{B_{1g}}^{(2)}=\frac{1}{\sqrt{2}}\left(\frac{\partial P_1}{\partial x_2}-\frac{\partial P_2}{\partial x_1}\right),\\
\left(\frac{\partial P}{\partial x}\right)_{B_{2g}}^{(1)}=\frac{1}{\sqrt{2}}\left(\frac{\partial P_1}{\partial x_3}+\frac{\partial P_3}{\partial x_1}\right),\quad \left(\frac{\partial P}{\partial x}\right)_{B_{2g}}^{(2)}=\frac{1}{\sqrt{2}}\left(\frac{\partial P_1}{\partial x_3}-\frac{\partial P_3}{\partial x_1}\right),\\
\left(\frac{\partial P}{\partial x}\right)_{B_{3g}}^{(1)}=\frac{1}{\sqrt{2}}\left(\frac{\partial P_2}{\partial x_3}+\frac{\partial P_3}{\partial x_2}\right),\quad \left(\frac{\partial P}{\partial x}\right)_{B_{3g}}^{(2)}=\frac{1}{\sqrt{2}}\left(\frac{\partial P_2}{\partial x_3}-\frac{\partial P_3}{\partial x_2}\right),
\end{gathered}
\qquad (\ .15)
$$

For strain, keeping just the symmetric representations, of the nine irreducible representations we end up with the six ones, $3A_{1g} + B_{1g} + B_{2g} + B_{3g}$. Correspondingly, from (B.4) it follows:

$$\begin{aligned} &A_{1g}: \quad u_A^{(1)} = u_{11}, \quad u_A^{(2)} = u_{22}, \quad u_A^{(3)} = u_{33}, \\ &B_{1g}: u_{1g} = \sqrt{2}u_{12};\; B_{2g}: u_{2g} = \sqrt{2}u_{13};\; B_{3g}: u_{3g} = \sqrt{2}u_{23} \end{aligned} \qquad (\ .16)$$

Components of the second-rank tensor $(\partial U/\partial x)_{\Gamma g}(\partial P/\partial x)_{\bar{\Gamma} \bar{g}}$ belong to the product $(3A_{1g} + B_{1g} + B_{2g} + B_{3g}) \times (3A_{1g} + 2B_{1g} + 2B_{2g} + 2B_{3g})$ where the first factor stands for $(\partial U/\partial x)_\Gamma$ and the second one corresponds to $(\partial P/\partial x)_\Gamma$. As mentioned above, of the $6 \times 9 = 54$ components of $(\partial U/\partial x)_{\Gamma g}(\partial P/\partial x)_{\bar{\Gamma} \bar{g}}$, we keep just totally-symmetric convolutions belonging to $A_{1g}$. From the product we can create nine scalars of the type $A_{1g} \times A_{1g}$ and six symmetric squares $[B_{1g}^2]$, $[B_{2g}^2]$, and $[B_{3g}^2]$, the total of fifteen scalar convolutions:

$$\begin{aligned} \Delta F = &- \sum_{m,n=1,2,3} f\left(A_{1g} \times A_{1g}\right)_{mn} u_{A_{1g}}^{(m)} \left(\frac{\partial P}{\partial x}\right)_{A_{1g}}^{(n)} \\ &- \sum_{n=1,2} \left\{ f\left[B_{1g}^2\right]_n u_{1g} \left(\frac{\partial P}{\partial x}\right)_{1g}^{(n)} + f\left[B_{2g}^2\right]_n u_{2g} \left(\frac{\partial P}{\partial x}\right)_{2g}^{(n)} + f\left[B_{3g}^2\right]_n u_{3g} \left(\frac{\partial P}{\partial x}\right)_{3g}^{(n)} \right\} \end{aligned} \qquad (\ .17)$$

Here the corresponding fifteen independent constants of flexoelectric coupling are introduced:

$$\begin{aligned} &f\left(A_{1g} \times A_{1g}\right)_{11} = f_{1111},\; f\left(A_{1g} \times A_{1g}\right)_{12} = f_{1122},\; f\left(A_{1g} \times A_{1g}\right)_{13} = f_{1133}, \\ &f\left(A_{1g} \times A_{1g}\right)_{21} = f_{2211},\; f\left(A_{1g} \times A_{1g}\right)_{22} = f_{2222},\; f\left(A_{1g} \times A_{1g}\right)_{23} = f_{2233}, \\ &f\left(A_{1g} \times A_{1g}\right)_{31} = f_{3311},\; f\left(A_{1g} \times A_{1g}\right)_{32} = f_{3322},\; f\left(A_{1g} \times A_{1g}\right)_{33} = f_{3333}, \\ &f\left[B_{1g}^2\right]_1 = \frac{1}{2}\left(f_{2112} + f_{1212} + 2f_{1122}\right),\; f\left[B_{1g}^2\right]_2 = \frac{1}{2}\left(f_{2112} + f_{1212} - 2f_{1122}\right), \\ &f\left[B_{2g}^2\right]_1 = \frac{1}{2}\left(f_{3113} + f_{1313} + 2f_{1133}\right),\; f\left[B_{2g}^2\right]_2 = \frac{1}{2}\left(f_{3113} + f_{1313} - 2f_{1133}\right), \\ &f\left[B_{3g}^2\right]_1 = \frac{1}{2}\left(f_{3223} + f_{2323} + 2f_{2233}\right),\; f\left[B_{3g}^2\right]_2 = \frac{1}{2}\left(f_{3223} + f_{2323} - 2f_{2233}\right) \end{aligned} \qquad (\ .18)$$

Including the hidden symmetry of Table II simplifies (B.18),

$$\begin{aligned} &f\left(A_{1g} \times A_{1g}\right)_{11} = f_{1111},\; f\left(A_{1g} \times A_{1g}\right)_{12} = f_{1122},\; f\left(A_{1g} \times A_{1g}\right)_{13} = f_{1133}, \\ &f\left(A_{1g} \times A_{1g}\right)_{21} = f_{2211},\; f\left(A_{1g} \times A_{1g}\right)_{22} = f_{2222},\; f\left(A_{1g} \times A_{1g}\right)_{23} = f_{2233}, \\ &f\left(A_{1g} \times A_{1g}\right)_{31} = f_{3311},\; f\left(A_{1g} \times A_{1g}\right)_{32} = f_{3322},\; f\left(A_{1g} \times A_{1g}\right)_{33} = f_{3333}, \\ &f\left[B_{1g}^2\right]_1 = f_{2211} + f_{1122},\; f\left[B_{1g}^2\right]_2 = f_{2211} - f_{1122}, \\ &f\left[B_{2g}^2\right]_1 = f_{3311} + f_{1133},\; f\left[B_{2g}^2\right]_2 = f_{3311} - f_{1133}, \\ &f\left[B_{3g}^2\right]_1 = f_{3322} + f_{2233},\; f\left[B_{3g}^2\right]_2 = f_{3322} - f_{2233} \end{aligned} \qquad (\ .19)$$

Here, as above, the fifteen constants of flexoelectric coupling are expressed in terms of just nine components, $f_{1111}$, $f_{2222}$, $f_{3333}$, $f_{1122}$, $f_{1133}$, $f_{2233}$, $f_{2211}$, $f_{3311}$, and $f_{3322}$. This means, with the hidden symmetry included, the eight coupling constants are not independent and can be expressed in terms of one another,

$$f\left[B_{1g}^2\right]_1 + f\left[B_{1g}^2\right]_2 = 2f\left(A_{1g} \times A_{1g}\right)_{21}, \quad f\left[B_{1g}^2\right]_1 - f\left[B_{1g}^2\right]_2 = 2f\left(A_{1g} \times A_{1g}\right)_{12}$$
$$f\left[B_{2g}^2\right]_1 + f\left[B_{2g}^2\right]_2 = 2f\left(A_{1g} \times A_{1g}\right)_{31}, \quad f\left[B_{2g}^2\right]_1 - f\left[B_{2g}^2\right]_2 = 2f\left(A_{1g} \times A_{1g}\right)_{13}$$
$$f\left[B_{3g}^2\right]_1 + f\left[B_{3g}^2\right]_2 = 2f\left(A_{1g} \times A_{1g}\right)_{32}, \quad f\left[B_{3g}^2\right]_1 - f\left[B_{3g}^2\right]_2 = 2f\left(A_{1g} \times A_{1g}\right)_{23}$$

Thus, the hidden symmetry results in the six constraints and, therefore, the number of independent coupling constants reduces from fifteen to just nine. Plugging (B.15), (B.16), and (B.19) into (B.17), we come to

$$\begin{aligned}\Delta F = &- f_{1111}u_{11}\frac{\partial P_1}{\partial x_1} - f_{2222}u_{22}\frac{\partial P_2}{\partial x_2} - f_{3333}u_{33}\frac{\partial P_3}{\partial x_3} \\ &- f_{1122}\left(u_{11}\frac{\partial P_2}{\partial x_2} + 2u_{12}\frac{\partial P_2}{\partial x_1}\right) - f_{2211}\left(u_{22}\frac{\partial P_1}{\partial x_1} + 2u_{12}\frac{\partial P_1}{\partial x_2}\right) \\ &- f_{1133}\left(u_{11}\frac{\partial P_3}{\partial x_3} + 2u_{13}\frac{\partial P_3}{\partial x_1}\right) - f_{3311}\left(u_{33}\frac{\partial P_1}{\partial x_1} + 2u_{13}\frac{\partial P_1}{\partial x_3}\right) \\ &- f_{2233}\left(u_{22}\frac{\partial P_3}{\partial x_3} + 2u_{23}\frac{\partial P_3}{\partial x_2}\right) - f_{3322}\left(u_{33}\frac{\partial P_2}{\partial x_2} + 2u_{23}\frac{\partial P_3}{\partial x_2}\right)\end{aligned} \qquad (\ .20)$$

**Appendix C. Effective flexoelectric response of the plate.**

The setup and interpretation of three-knife load experiment to determine flexoelectric coefficients

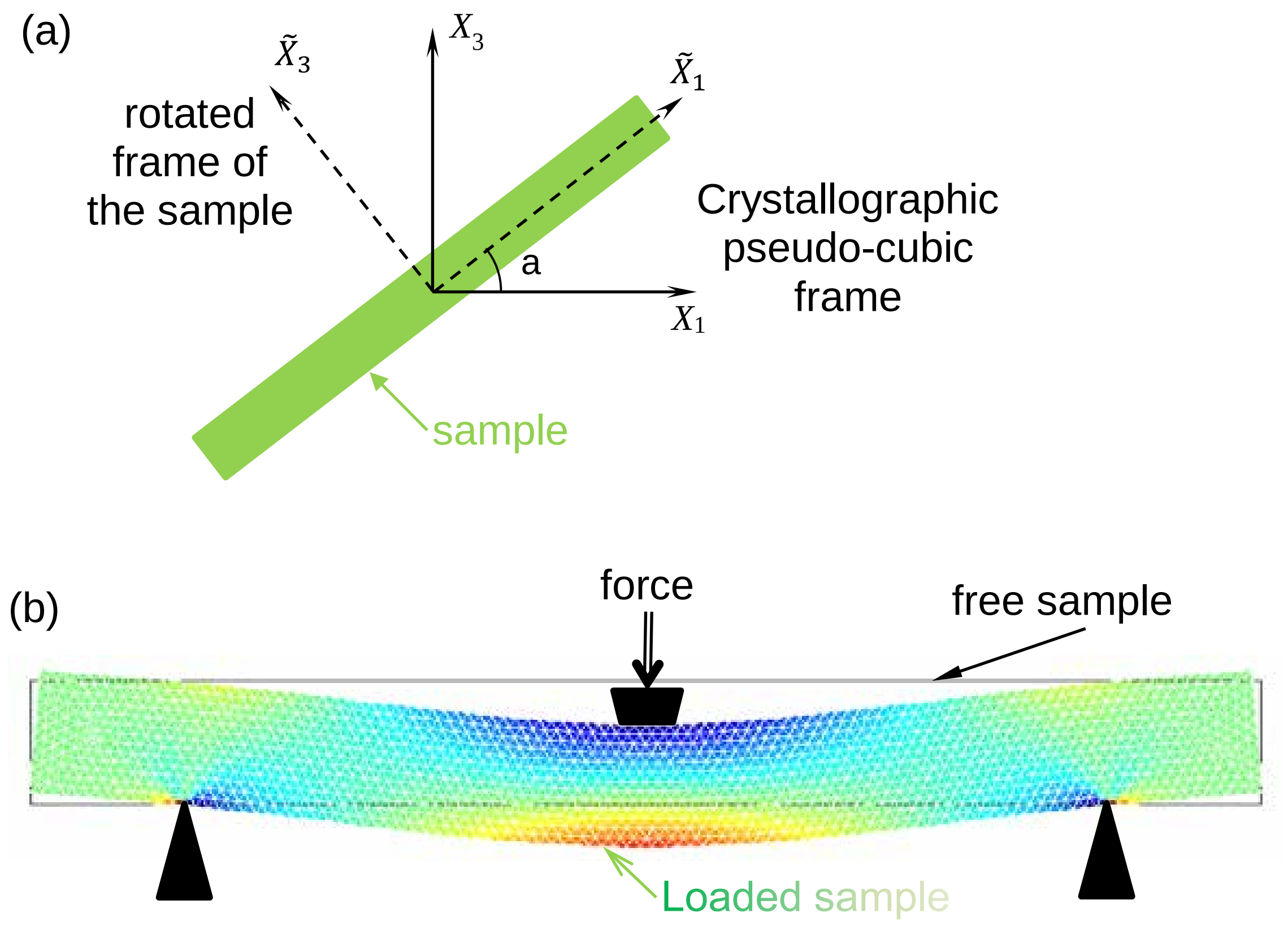

Figure C1. (a) Coordinate transformation from crystallographic (pseudo-cubic) frame $X_1, X_2, X_3$ by the rotation around $X_2$ axis on angle a to sample-related frame $\tilde{X}_1, \tilde{X}_2, \tilde{X}_3$. (b) Typical three-knife experiment of plate bending, black solid lines represent cross-section of the initial, rainbow-colored shape represents the loaded sample with aggregated strain.

The coordinate frame transformation is (see also Fig. C1a)

$$\tilde{X}_1 = \cos(\alpha)X_1 + \sin(\alpha)X_3$$

$$\tilde{X}_3 = -\sin(\alpha)X_1 + \cos(\alpha)X_3 \qquad \text{(C.1)}$$

For 4/mmm point symmetry group transformation yields

$$\tilde{f}_{1133} = f_{1133} + \sin^2(\alpha)\begin{pmatrix} f_{3311} - f_{1133} + \\ \cos^2(\alpha)\left(f_{1111} - f_{1133} + f_{3333} - f_{3311} - 2(f_{1313} + f_{1331})\right) \end{pmatrix} \qquad \text{(C.2a)}$$

$$\tilde{f}_{2233} = f_{1122} + \cos^2(\alpha)(f_{1133} - f_{1122}) \qquad \text{(C.2b)}$$

$$\tilde{f}_{1333} = \cos(\alpha)\sin(\alpha)\begin{pmatrix} +\cos^2(\alpha)(f_{3333} - f_{1133} - f_{1313} - f_{1331}) \\ -\sin^2(\alpha)(f_{1111} - f_{3311} - f_{1313} - f_{1331}) \end{pmatrix} \qquad \text{(C.2c)}$$

$$\tilde{f}_{3333} = f_{3333} + \sin^2(\alpha)\begin{pmatrix} f_{1111} - f_{3333} - \\ \cos^2(\alpha)\left(f_{1111} - f_{1133} + f_{3333} - f_{3311} - 2(f_{1313} + f_{1331})\right) \end{pmatrix} \qquad \text{(C.2d)}$$

For m3m point symmetry group transformation yields

$$\tilde{f}_{1133} = f_{1122} + 2\sin^2(\alpha)\cos^2(\alpha)\left(f_{1111} - f_{1122} - (f_{1212} + f_{1221})\right) \qquad \text{(C.3a)}$$

$$\tilde{f}_{1313} = f_{1212} - \sin^2(\alpha)(f_{1212} - f_{1221}) + 2\sin^2(\alpha)\cos^2(\alpha)\big(f_{1111} - f_{1122} - (f_{1212} + f_{1221})\big) \tag{C.3b}$$

$$\tilde{f}_{1331} = f_{1221} - \sin^2(\alpha)(f_{1221} - f_{1212}) + 2\sin^2(\alpha)\cos^2(\alpha)\big(f_{1111} - f_{1122} - (f_{1212} + f_{1221})\big) \tag{C.3c}$$

$$\tilde{f}_{2233} = f_{1122} \tag{C.3d}$$

$$\tilde{f}_{1333} = \cos(\alpha)\sin(\alpha)\big(\cos^2(\alpha) - \sin^2(\alpha)\big)(f_{1111} - f_{1122} - f_{1212} - f_{1221}) \tag{C.3e}$$

$$\tilde{f}_{3333} = f_{1111} - 2\sin^2(\alpha)\cos^2(\alpha)\big(f_{1111} - f_{1122} - (f_{1212} + f_{1221})\big) \tag{C.3f}$$

Below we consider the bending problem for a plate (see Fig. C1b), having developed surface rotated on angle $\alpha$ from principal (pseudo-) cubic direction (see Fig. C1a). Below we omit the tilde sings for clarity and consider only normal component of polarization, $P_3$

Equations of state could be obtained after the variation of the corresponding thermodynamic potential in the following form

$$\sigma_{ij} = c_{ijkl}u_{kl} + f_{ijkl}P_{k,l} \tag{C.4a}$$

$$\alpha_3 P_3 - g_{3333}P_{3,33} - f_{ij3l}u_{ij,l} = E_3 \tag{C.4b}$$

Using compatibility conditions and considering the case of strain field, depending only on $X_3$, one could get the following restrictions on the strain tensor components:

$$u_{11,33} = u_{12,33} = u_{22,33} = 0 \tag{C.5}$$

while other components could have an arbitrary dependence on $X_3$. The relations (C.5a) means that three components of strain tensor have linear dependences

$$u_{11} = u_{11}^{(0)} + \frac{X_3}{R_1} \tag{C.6a}$$

$$u_{12} = u_{12}^{(0)} \tag{C.6b}$$

$$u_{22} = u_{22}^{(0)} + \frac{X_3}{R_2} \tag{C.6c}$$

Here constants, $u_{11}^{(0)}, u_{11}^{(0)}, u_{11}^{(0)}$, $R_1$ and $R_2$ should be determined from boundary conditions at side faces of plate (usually in Saint-Venant approximation). However, as far as we are interested in the strain gradient estimation in order to get expression for polarization from (C.4b), "an exact" expressions for constants from Eqs.(C.6) are not necessary, since radii of plate curvature $R_1$ and $R_2$ could be estimated from the shape of the stressed plate. Next we recall boundary conditions for stresses at the developed surface of the plate, namely $\sigma_{ij}n_j\big|_S = 0$, and condition of mechanical equilibrium, $\sigma_{ij,j} = 0$. For the case of 1D dependences these two relations give the following

$$\sigma_{13} = \sigma_{23} = \sigma_{33} = 0 \tag{C.7}$$

Finally, taking into account Eqs.(C.2), (C.3), and (C.6), one could get the evident form of equations of state, $\sigma_{13} = 2c_{1313}u_{13} + c_{1333}u_{33} + f_{1333}P_{3,3}$ and $\sigma_{33} = c_{1133}u_{11} + 2c_{1333}u_{13} + c_{2233}u_{22} + c_{3333}u_{33} + f_{3333}P_{3,3}$. so that relations (C.7) give the system of equations for the unknown strains $u_{13}$ and $u_{33}$:

$$2c_{1313}u_{13} + c_{1333}u_{33} = -f_{1333}P_{3,3} \tag{C.8a}$$

$$2c_{1333}u_{13} + c_{3333}u_{33} = -c_{1133}u_{11} - c_{2233}u_{22} - f_{3333}P_{3,3} \tag{C.8b}$$

The solution of (C.8) is

$$2u_{13} = \frac{+c_{1333}(c_{1133}u_{11} + c_{2233}u_{22} + f_{3333}P_{3,3}) - c_{3333}f_{1333}P_{3,3}}{c_{1313}c_{3333} - c_{1333}^2} \tag{C.9a}$$

$$u_{33} = \frac{-c_{1313}(c_{1133}u_{11} + c_{2233}u_{22} + f_{3333}P_{3,3}) + c_{1333}f_{1333}P_{3,3}}{c_{1313}c_{3333} - c_{1333}^2} \tag{C.9b}$$

Next, using evident form of equation of state (C.4b), $\alpha_3 P_3 - g_{3333}P_{3,33} - f_{1133}u_{11,3} - f_{2233}u_{22,3} - f_{3333}u_{33,3} - 2f_{1333}u_{13,3} = E_3$, one could get the following equation for polarization inside the slab:

$$\alpha_3 P_3 - \left(g_{3333} - \frac{c_{1313} f_{3333}^2 + c_{3333} f_{1333}^2 - 2c_{1333} f_{1333} f_{3333}}{c_{1313} c_{3333} - c_{1333}^2}\right) P_{3,33} = E_3 +$$

$$+ \left(f_{1133} - \frac{(f_{3333} c_{1313} - f_{1333} c_{1333}) c_{1133}}{c_{1313} c_{3333} - c_{1333}^2}\right) u_{11,3} + \left(f_{2233} - \frac{(f_{3333} c_{1313} - f_{1333} c_{1333}) c_{2233}}{c_{1313} c_{3333} - c_{1333}^2}\right) u_{22,3} \quad \text{(C.10)}$$

One could see that according to Eqs. (C.6) there is a uniform gradient of strain leading to a sort of constant “flexoelectric field” in the right-hand side of Eq. (C.10)

$$E_3^{(flexo)} = \left(f_{1133} - \frac{(f_{3333} c_{1313} - f_{1333} c_{1333}) c_{1133}}{c_{1313} c_{3333} - c_{1333}^2}\right) \frac{1}{R_1} + \left(f_{2233} - \frac{(f_{3333} c_{1313} - f_{1333} c_{1333}) c_{2233}}{c_{1313} c_{3333} - c_{1333}^2}\right) \frac{1}{R_2} \quad \text{(C.11)}$$

Two terms from (C.11) represent the flexoelectric response to plate bending into two perpendicular directions, leading to plate transforming to “cup” like shape with two main values of curvature, $1/R_1$ and $1/R_2$. Hence we could introduce coefficients of flexoelectric response in the following form:

$$f_{13}^{eff} = f_{1133} - \frac{c_{1133}}{c_{1313} c_{3333} - c_{1333}^2} (f_{3333} c_{1313} - f_{1333} c_{1333}) \quad \text{(C.12a)}$$

$$f_{23}^{eff} = f_{2233} - \frac{c_{2233}}{c_{1313} c_{3333} - c_{1333}^2} (f_{3333} c_{1313} - f_{1333} c_{1333}) \quad \text{(C.12b)}$$